# Robust Total Least Mean M-Estimate normalized subband filter Adaptive Algorithm for impulsive noises and noisy inputs

Haiquan Zhao, *Senior Member*, *IEEE*, Zian Cao, and Yida Chen

***Abstract*—When the input signal is correlated input signals, and the input and output signal is contaminated by Gaussian noise, the total least squares normalized subband adaptive filter (TLS-NSAF) algorithm shows good performance. However, when it is disturbed by impulse noise, the TLS-NSAF algorithm shows the rapidly deteriorating convergence performance. To solve this problem, this paper proposed the robust total minimum mean M-estimator normalized subband filter (TLMM-NSAF) algorithm. In addition, this paper also conducts a detailed theoretical performance analysis of the TLMM-NSAF algorithm and obtains the stable step size range and theoretical steady-state mean square error of the algorithm. To further improve the performance of the algorithm, we also propose a new variable step size (VSS) method of the algorithm. Finally, the robustness of our proposed algorithm and the consistency of theoretical and simulated values are verified by computer simulations of system identification and echo cancellation under different noise models.**

***Index Terms*—M-estimate, errors-in-variables model, total least squares, subband adaptive filter, correlated input.**

## I. Introduction

Adaptive filtering has been widely used in echo cancellation, system identification and other fields [1]. The classical normalized least mean square (NLMS) algorithm is universally adopted due to its efficiency and simplicity [2]. However, when the input signal is a correlated input signal, the convergence of the NLMS algorithm becomes poor. The Affine Projection Algorithm (APA) can solve this problem [3], but it is computationally expensive [4]. In [7], Lee et al. proposed the normalized subband adaptive filter (NSAF) algorithm, which converges better than the NLMS algorithm under correlated inputs without significantly increasing the computational complexity.

However, the above traditional adaptive filtering algorithms only assume that the output noise is polluted. Under the EIV model where both the input and output signals have noise interference, the convergence performance of the traditional adaptive filtering algorithm will be degraded. In the past, the bias-compensated method was often used under the EIV model, which introduced bias compensation vectors to compensate for estimation errors caused by noise in the input signal. Many adaptive filtering algorithms utilizing the bias-compensated method are already proposed [14]-[16].

Different from bias-compensated method, The total least squares (TLS) has good convergence performance when dealing with Errors-in-variables (EIV) models. Reference [9] extended the TLS algorithm to subbands and proposed the TLS-NSAF algorithm, which showed good performance under the EIV model. Unfortunately, the performance of the TLS-NSAF algorithm deteriorates severely when disturbed by impulse noise. Although the impulsive noise is short in duration, it has a large amplitude and causes the algorithm's convergence behavior to fluctuate wildly. To acquire the stable convergence in impulsive noise, the sign algorithm was firstly proposed by minimizing the absolute value of error signal, but with slow convergence rate [33][34]. In addition to sign algorithm, The M-estimator function (MF) has the ability to distinguish outliers,and excellent performance against impulse noise, so it has been used to develop robust adaptive algorithms against impulsive noise [10]-[12],[23].

In this paper, in order to solve the problem that the performance of TLS-NSAF algorithm is seriously degraded when it suffers from impulse noise, the robust total least mean M-Estimate normalized subband filter (TLMM-NSAF) algorithm is proposed. The algorithm shows good performance under the EIV model and is robust to impulse noise. More importantly, a detailed theoretical analysis is provided for the TLMM-NSAF algorithm. In order to further improve the convergence speed of the algorithm and reduce the steady-state error of the algorithm， we used the existing VSS strategy[18][29][35] to improve the traditional convex combination method [30], and proposed a new VSS method. Finally, the simulations of system identification and echo cancellation under different noise models verify the robustness of our proposed algorithm and the consistency of theoretical and simulated values. The main contributions of this paper are as follows.

1) The proposed TLMM-NSAF algorithm solves the problem that the performance of the traditional subband adaptive filtering algorithm is seriously degraded under the

This work was partially supported by National Natural Science Foundation of China (grant: 62171388, 61871461, 61571374), Fundamental Research Funds for the Central Universities (grant: 2682021ZTPY091).

Haiquan Zhao, Zian Cao and Yida Chen are with the Key Laboratory of Magnetic Suspension Technology and Maglev Vehicle, Ministry of Education, and the School of Electrical Engineering, Southwest Jiaotong University, Chengdu, 610031, China. (e-mail: hqzhao_swjtu@126.com; ; ziancao_swjtu@126.com;yidachen_swjtu@126.com).

Corresponding author: Haiquan Zhao.

EIV model and after being interfered by impulsive noise.

2) The local stability, computational complexity and steady-state MSD of the TLMM-NSAF algorithm under impulsive noise are analyzed. The analytical results are supported by simulations.

3) In order to improve the performance of the TLMM-NSAF algorithm, the existing variable step size strategy is improved, and a Variable step convex combination total least squares normalized subband adaptive filter algorithm (VSS-CTLMM-NSAF) is proposed.

4) The proposed algorithm is compared with other algorithms under the simulation of system identification and acoustic echo cancellation, and the simulation results verify the superiority of the proposed algorithm.

The remaining part of this paper is organized as follows. Section II introduces the subband structure of the input signal contaminated by noise and reviews the EIV model and the TLS-NSAF algorithm. Section III deduces the TLMM-NSAF algorithm in detail. Section IV analyzes the local stability of the algorithm under impulse noise interference, and obtains the step size range to ensure the stability of the algorithm. Section V calculates the steady-state mean squared deviation of the TLMM-NSAF algorithm. Section VI proposed a new variable step size method. Section VII analyzes the computational complexity of the algorithm. Section VIII presents the simulation results. The last section gives the conclusion.

## II. BRIEF REVIEW

### A. EIV model and subband adaptive filter (SAF) structure

There exists a linear system which is described as:

$$d(\mathrm{n}) = \boldsymbol{h}^T \boldsymbol{x}(n), \tag{1}$$

where $\boldsymbol{h}\in\mathbb{R}^{L\times1}$ is unknown weight vector to be found, $\boldsymbol{x}(n)=[x(n),\ldots,x(n-L+1)]^T\in\mathbb{R}^{L\times1}$ and $d(n)\in\mathbb{R}$ are input vector and corresponding output signal respectively.

In the EIV model, both the input and desired signals are disturbed by noise, and can be expressed as:

$$\tilde{\boldsymbol{x}}(n) = \boldsymbol{x}(n) + \boldsymbol{u}(n), \tag{2}$$

$$\tilde{d}(n) = d(n) + v(n), \tag{3}$$

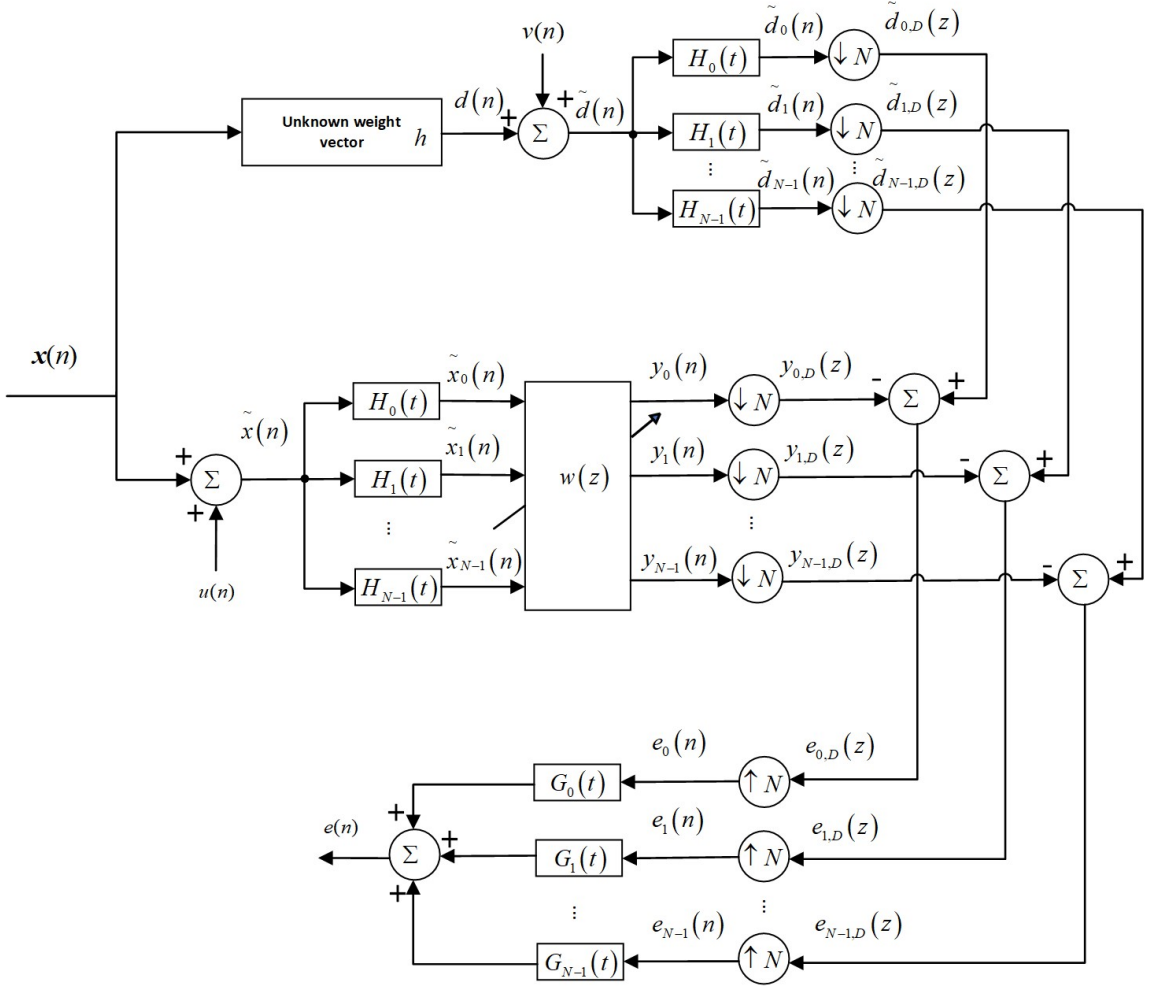

Fig. 1. Subband adaptive filter under EIV model

where $\boldsymbol{u}(n)\in\mathbb{R}^{L\times1}$ and $v(n)\in\mathbb{R}$ are the input and output noise signals, of which the variance is $\sigma_{in}^2$ and $\sigma_o^2$, respectively.

The subband adaptive filter structure under the EIV model is given by Fig. 1: The analysis filter $H_i(t), i=0,\ldots,N-1$ decomposes the input signal $\tilde{\boldsymbol{x}}_i(n)$ and the desired signal $\tilde{d}_i(n)$ proportionally to obtain the subband input signal $\tilde{\boldsymbol{x}}_i(n)$ and the subband desired signal $\tilde{d}_i(n)$. $y_i(n)$ is the subband output signal. Strict downsampling of subband signals $\tilde{d}_i(n)$ and $y_i(n)$ can generate signals $\tilde{d}_{i,D}(z)$ and $y_{i,D}(z)$ with lower sampling rates. The subband error signal is obtained by

$$e_{i,D}(z) = \tilde{d}_{i,D}(z) - y_{i,D}(z) = \tilde{d}_{i,D}(z) - \tilde{\boldsymbol{x}}_i^T(z)\boldsymbol{w}(z), \tag{4}$$

where the subband input signal and the weight vector are expressed as $\tilde{\boldsymbol{x}}_i(z)=[\tilde{x}_i(zN),\tilde{x}_i(zN-1),\ldots\tilde{x}_i(zN-L+1)]^T\in\mathbb{R}^{L\times1}$, and $\boldsymbol{w}(z)=[w_0(z),w_1(z),\ldots,w_{L-1}(z)]^T\in\mathbb{R}^{L\times1}$. The index of the decimated sequence is defined as variable $z$ in this paper.

### B. TLS-NSAF algorithm

In [9], the author applies the TLS method to subband adaptive filtering and proposes the TLS-NSAF algorithm.To minimize the interference of input noise and output noise to the system, the TLS method in a single subband can be described as the following minimization problem [17]

$$\min_{w} \boldsymbol{J}_{i,z}(\boldsymbol{w}) = \sum_{j=1}^{z} \frac{\left(\tilde{d}_{i,D}(j) - \tilde{\boldsymbol{x}}_i^T(j)\boldsymbol{w}\right)^2}{\|\boldsymbol{w}\|^2 + \theta_i} \tag{5}$$

where $\theta_i = \sigma_{i,o}^2 / \sigma_{i,in}^2$ is the ratio of the output noise variance to the input signal noise variance, and in practice the values of $\sigma_{in}^2$ and $\sigma_o^2$ are often very close or the same [19].

Because the input data and the input and output noises are correlation ergodic processes, the time average in (5) is replaced with the expectation and obtain the following alternative cost function:

$$J_i(\boldsymbol{w}) = E\left[\frac{\left(\tilde{d}_{i,D}(z) - \tilde{\boldsymbol{x}}_i^T(z)\boldsymbol{w}\right)^2}{\|\boldsymbol{w}\|^2 + \theta_i}\right] = E\left[\frac{e_{i,D}^2(z)}{\|\bar{\boldsymbol{w}}\|^2}\right], \tag{6}$$

where $\bar{\boldsymbol{w}} = [\boldsymbol{w}^T, -\sqrt{\theta_i}]^T$ . Then, the cost function in the full-band filter is obtained as

$$J_{TLS-SAF}(\boldsymbol{w}) = \sum_{i=0}^{N-1} E\left[\frac{e_{i,D}^2(z)}{\|\bar{\boldsymbol{w}}\|^2}\right] \tag{7}$$

Adding the normalization term to $J_{TLS-SAF}(\boldsymbol{w})$ in (9) because of the advantages of the multi-band-structured SAF (MSAF) [7], [8], [21]. The cost function of the TLS-NSAF algorithm can be obtained as

$$\boldsymbol{J}_{TLS-NSAF}(\boldsymbol{w}) = \sum_{i=0}^{N-1} E\left[\frac{e_{i,D}^2(z)}{\|\bar{\boldsymbol{w}}\|^2 \|\tilde{\boldsymbol{x}}_i(z)\|^2}\right]. \tag{8}$$

## III. THE PROPOSED TLMM ALGORITHM

Inspired by the least mean M-estimate LMM algorithm [24], this section propose a new robust adaptive algorithm called total least mean M-estimate normalized subband adaptive filtering (TLMM-NSAF) algorithm, which is derived based on the following cost function

$$\boldsymbol{J}_{TLMM-NSAF}(\boldsymbol{w})=\sum_{i=0}^{N-1}E\left[\frac{\rho\left(e_{i,D}(z)\right)}{\|\bar{\boldsymbol{w}}\|^2\|\tilde{\boldsymbol{x}}_i(z)\|^2}\right]. \tag{9}$$

where $\rho(.)$ is the M-estimate function. The M-estimate function is also used in the some existing literatures [24-26], [20]. The modified Huber function is used as the M-estimate function due to its simplicity and efficiency

$$\rho\left(e_{i,D}(z)\right)=\begin{cases}\dfrac{e_{i,D}^2(z)}{2}, & \left|e_{i,D}(z)\right|<\xi_i\\ \dfrac{\xi^2\|\bar{\boldsymbol{w}}\|^2}{2}, & \left|e_{i,D}(z)\right|\geq\xi_i\end{cases} \tag{10}$$

where $\xi_i$ is the threshold parameter for controlling the suppression of impulsive noise. In general, $\xi_i$ is adjusted by $\xi_i=2.576\hat{\sigma}_{e,i}(n)$ [25]. The robust estimator $\hat{\sigma}_{e,i}(n)$ can be obtained by

$$\hat{\sigma}_{e,i}^2(z)=\lambda_\sigma\hat{\sigma}_{e,i}^2(z-1)+k(1-\lambda_\sigma)med\left(A_{e,i}(z)\right)$$

where $\lambda_\sigma$ is a forgrtting factor and $0<\lambda_\sigma<1$. $k=1.483\left(1+5/\left(N_w-1\right)\right)$ is the correction factor. The $N_w$ is usually chosen between 5 and 9 [26]. $med(.)$ is the median operator and $A_{e,i}(z)=\left[e_{i,D}^2(z),e_{i,D}^2(z-1),...,e_{i,D}^2(z-N_w+1)\right]$.

Using the gradient descent method, the gradient vector $\boldsymbol{g}_{TLMM-NSAF}(\boldsymbol{w})$ of $\boldsymbol{J}_{TLMM-NSAF}(\boldsymbol{w})$ can be calculated as:

$$\boldsymbol{g}_{TLMM-NSAF}(\boldsymbol{w})=\frac{\partial J_{TLMM-NSAF}}{\partial w}$$
$$=\begin{cases}-E\left[\displaystyle\sum_{i=0}^{N-1}\frac{\|\bar{\boldsymbol{w}}\|^2e_{i,D}(z)\tilde{\boldsymbol{x}}_i(z)+e_{i,D}^2(z)\boldsymbol{w}}{\|\bar{\boldsymbol{w}}\|^4\|\tilde{\boldsymbol{x}}_i(z)\|^2}\right] & \left|e_{i,D}(z)\right|<\xi_i\\ 0 & \left|e_{i,D}(z)\right|\geq\xi_i\end{cases} \tag{11}$$

Because the subband input power has the chi-square distribution under $L$ degrees of freedom [13], (11) can be simplified as:

$$\boldsymbol{g}_{TLMM-NSAF}(\boldsymbol{w})=\frac{\partial J_{TLMM-NSAF}}{\partial w}$$
$$=\begin{cases}-E\left[\displaystyle\sum_{i=0}^{N-1}\frac{\|\bar{\boldsymbol{w}}\|^2e_{i,D}(z)\tilde{\boldsymbol{x}}_i(z)+e_{i,D}^2(z)\boldsymbol{w}}{\|\bar{\boldsymbol{w}}\|^4(L-2)\sigma_{i,\tilde{x}}^2}\right] & \left|e_{i,D}(z)\right|<\xi_i\\ 0 & \left|e_{i,D}(z)\right|\geq\xi_i\end{cases} \tag{12}$$

After replacing the expected value with the instantaneous value, the instantaneous gradient vector can be further obtained:

$$\hat{\boldsymbol{g}}_{TLMM-NSAF}(\boldsymbol{w})$$
$$=\begin{cases}-\displaystyle\sum_{i=0}^{N-1}\frac{\|\bar{\boldsymbol{w}}\|^2e_{i,D}(z)\tilde{\boldsymbol{x}}_i(z)+e_{i,D}^2(z)\boldsymbol{w}}{\|\bar{\boldsymbol{w}}\|^4(L-2)\sigma_{i,\tilde{x}}^2} & \left|e_{i,D}(z)\right|<\xi_i\\ 0 & \left|e_{i,D}(z)\right|\geq\xi_i\end{cases} \tag{13}$$

Using (13), can obtain the weight vector update formula of the TLMM-NSAF algorithm as:

$$\boldsymbol{w}(z+1)=\boldsymbol{w}(z)-\mu\hat{\boldsymbol{g}}_{TLMM-NSAF}(\boldsymbol{w}(z))$$
$$=\begin{cases}\boldsymbol{w}(z)+\mu\displaystyle\sum_{i=0}^{N-1}\frac{\|\bar{\boldsymbol{w}}(z)\|^2e_{i,D}(z)\tilde{\boldsymbol{x}}_i(z)+e_{i,D}^2(z)\boldsymbol{w}(z)}{\|\bar{\boldsymbol{w}}(z)\|^4(L-2)\sigma_{i,\tilde{x}}^2} & \left|e_{i,D}(z)\right|<\xi_i\\ \boldsymbol{w}(z) & \left|e_{i,D}(z)\right|\geq\xi_i\end{cases} \tag{14}$$

It can be proved from (13) that the output error will be large due to the impulse noise, which will lead to $\left|e_{i,D}(z)\right|\geq\xi_i$, and the weight vector will not be updated at this time, thus effectively avoiding the interference of impulse noise.

The pseudocode of the proposed TLMM-NSAF algorithm is summarized in TABLE I.

TABLE I
THE PSEUDOCODE OF THE MMC-NSAF ALGORITHM.

**Initialization:** $\mathrm{w}(0)=\mathbf{0}$

**Parameters:** $\mu$, $N_w$, $\lambda_\sigma$, $k=1.483\left(1+5/\left(N_w-1\right)\right)$

**For** $z$= 0,1,2, ...

  **For** i=[0,N-1]

  $e_{i,D}(z)=\tilde{d}_{i,D}(z)-\tilde{\boldsymbol{x}}_i^T(z)\boldsymbol{w}(z)$

  $A_{e,i}(z)=\left[e_{i,D}^2(z),e_{i,D}^2(z-1),...,e_{i,D}^2(z-N_w+1)\right]$

  $\hat{\sigma}_{e,i}^2(z)=\lambda_\sigma\hat{\sigma}_{e,i}^2(z-1)+k(1-\lambda_\sigma)med\left(A_{e,i}(z)\right)$

  $\xi_i=2.576\hat{\sigma}_{e,i}(n)$

  **If** $\left|e_{i,D}(z)\right|<\xi_i$

  $\boldsymbol{w}(z+1)=\boldsymbol{w}(z)+\mu\sum_{i=0}^{N-1}\frac{\|\bar{\boldsymbol{w}}(z)\|^2e_{i,D}(z)\tilde{\boldsymbol{x}}_i(z)+e_{i,D}^2(z)\boldsymbol{w}(z)}{\|\bar{\boldsymbol{w}}(z)\|^4(L-2)\sigma_{i,\tilde{x}}^2}$

  **else**

  $\boldsymbol{w}(z+1)=\boldsymbol{w}(z)$

  **End**

  **End**

**End**

## IV. LOCAL STABILITY ANALYSIS

This section analyzes the local extreme points of the cost function $\boldsymbol{J}_{TLMM-NSAF}(\boldsymbol{w})$ and analyze the local stability of the TLMM-NSAF algorithm, and determine the range of step sizes that the algorithm can stably converge.

To make the analysis tractable, the following assumptions are provided, which are common assumptions in adaptive filtering analysis [1].

*Assumption 1*: The input noise vector $\boldsymbol{u}_i(z)$, the weight vector $\boldsymbol{w}(z)$, the input vector $\boldsymbol{x}_i(z)$ and the output noise $v_i(z)$ are independent mutually [1], [14], [15].

*Assumption 2*: The analysis filter bank is paraunitary [13], [15].

*Assumption 3*: When $L$ is long, the fluctuation of $\|\tilde{\boldsymbol{x}}_i(z)\|^2$ from one time to the next is small enough [6], [13].

*A. Local extreme point*

It can be seen from equations (12)-(14) that when $|e_{i,D}(z)| \geq \xi_i$, the gradient vector $\hat{\boldsymbol{g}}_{TLMM-NSAF}(\boldsymbol{w}) = 0$, and the weight vector will not be updated at this time. Therefore, in the following theoretical analysis, only the situation of $|e_{i,D}(z)| < \xi_i$ needs to be considered.

From formula (1), $\boldsymbol{h}$ is the ideal optimal weight vector and the value of the gradient vector $\hat{\boldsymbol{g}}_{TLMM-NSAF}(\boldsymbol{w})$ in (13) at $\boldsymbol{h}$ can be written as

$$\boldsymbol{g}_{TLMM-NSAF}(\boldsymbol{h}) = -\sum_{i=0}^{N-1} \frac{\|\bar{\boldsymbol{h}}\|^2 E[e_{i,D}(z)\tilde{\boldsymbol{x}}_i(z)] + E[e_{i,D}^2(z)]\boldsymbol{h}}{\|\bar{\boldsymbol{h}}\|^4 (L-2)\sigma_{i,\tilde{x}}^2} \quad (15)$$

The weight vector update formula in (14) also can be written as

$$\boldsymbol{w}(z+1) = \boldsymbol{w}(z) + \mu \sum_{i=0}^{N-1} \frac{\|\bar{\boldsymbol{w}}(z)\|^2 e_{i,D}(z)\tilde{\boldsymbol{x}}_i(z) + e_{i,D}^2(z)\boldsymbol{w}(z)}{\|\bar{\boldsymbol{w}}(z)\|^4 (L-2)\sigma_{i,\tilde{x}}^2}.$$

The error $e_{i,D}(z)$ at $\boldsymbol{h}$ can be obtained by

$$e_{i,D}(z) = \tilde{d}_{i,D}(z) - \boldsymbol{h}^T \tilde{\boldsymbol{x}}_i(z) = v_i(z) - \boldsymbol{h}^T \boldsymbol{u}_i(z), \quad (16)$$

Based on Assumption 1, the expectation in equation (15) can be calculated as:

$$E[e_{i,D}^2(z)] = E[v_i^2(z)] + E[(\boldsymbol{h}^T \boldsymbol{u}_i(z))^2] = \|\bar{\boldsymbol{h}}\|^2 \sigma_{i,in}^2, \quad (17)$$

$$E[e_{i,D}(z)\tilde{\boldsymbol{x}}_i(z)] = E[(v_i(z) - \boldsymbol{h}^T \boldsymbol{u}_i(z))(\boldsymbol{x}_i(z) + \boldsymbol{u}_i(z))] = -\boldsymbol{h}\sigma_{i,in}^2. \quad (18)$$

Substituting (17) and (18) to (15), we can calculate $\boldsymbol{g}_{TLMM-NSAF}(\boldsymbol{h}) = 0$. So, $\boldsymbol{h}$ is the critical point of $J_{TLMM-NSAF}(\boldsymbol{w})$. To further prove that $\boldsymbol{h}$ is a local extreme point, it is necessary to calculate the Hessian matrix $\boldsymbol{H}_{TLMM-NSAF}(\boldsymbol{w})$ of $J_{TLS-NSAF}(\boldsymbol{w})$.

By differentiating (12) with respect to $\boldsymbol{w}^T$, the Hessian matrix can be calculated as

$$\boldsymbol{H}_{TLMM-NSAF}(\boldsymbol{w}) = \frac{\partial \boldsymbol{g}_{TLMM-NSAF}(\boldsymbol{w})}{\partial \boldsymbol{w}^T} = \frac{1}{\|\bar{\boldsymbol{w}}\|^2}\left[\sum_{i=0}^{N-1} \frac{E\left[\tilde{\boldsymbol{x}}_i(z)\tilde{\boldsymbol{x}}_i^T(z)\right]}{(L-2)\sigma_{i,\tilde{x}}^2}\boldsymbol{I} - \sum_{i=0}^{N-1} \frac{E[e_{i,D}^2(z)]}{\|\bar{\boldsymbol{w}}\|^2 (L-2)\sigma_{i,\tilde{x}}^2}\boldsymbol{I} - \boldsymbol{g}_{TLMM-NSAF}(\boldsymbol{w})\boldsymbol{w}^T - \boldsymbol{w}\boldsymbol{g}_{TLMM-NSAF}^T(\boldsymbol{w})\right]. \quad (19)$$

Then the Hessian matrix at $\boldsymbol{h}$ is expressed as :

$$\boldsymbol{H}_{TLMM-NSAF}(\boldsymbol{h}) = \frac{1}{\|\bar{\boldsymbol{h}}\|^2}\left[\sum_{i=0}^{N-1} \gamma_i \boldsymbol{I} - \sum_{i=0}^{N-1} \frac{\sigma_{i,in}^2 \boldsymbol{I}}{(L-2)\sigma_{i,\tilde{x}}^2}\right], \quad (20)$$

where $\gamma_i \boldsymbol{I} \approx \frac{E\left[\tilde{\boldsymbol{x}}_i(z)\tilde{\boldsymbol{x}}_i^T(z)\right]}{E\left[\|\tilde{\boldsymbol{x}}_i(z)\|^2\right]}$ and $\gamma_i$ is only influenced by the correlatedness of the input signal and affects the convergence rate. When the number of subbands is sufficiently large or the original input signal is the white signal, the subband input signal is a white signal[9]. In this case $\gamma_i \approx \frac{1}{L}$.When the input signal is the relevant input signal, $\gamma_i$ is determined as the average value of $\alpha_i$ and $\frac{1}{L}$ for the correlated input signal. Where $\alpha_i$ means the minimum eigenvalue of the covariance matrix of the $i$ th subband input. The content of $\gamma_i$ is covered in detail in [27].

In a noisy subband input signal, the variance of the input noise is smaller than the variance of the input signal, so it can be obtained:

$$\sigma_{i,\tilde{x}}^2 > \sigma_{i,in}^2. \quad (21)$$

To prove that $\boldsymbol{h}$ is a local minimum point, it is also necessary to prove that the Hessian matrix $\boldsymbol{H}_{TLMM-NSAF}(\boldsymbol{w})$ is positive definite. According to (20), when the following conditions are satisfied, the $\boldsymbol{H}_{TLMM-NSAF}(\boldsymbol{w})$ is positive definite.

$$(L-2)\frac{\sigma_{i,\tilde{x}}^2}{\sigma_{i,in}^2} > \frac{1}{\gamma_i}. \quad (22)$$

Because of the filter order L is much larger than 2 and $\frac{\sigma_{i,\tilde{x}}^2}{\sigma_{i,in}^2} > 1$ can be known from equation (21), so the condition of equation (22) is satisfied and it can be proved that the $\boldsymbol{H}_{TLMM-NSAF}(\boldsymbol{w})$ is a positive definite matrix, and $\boldsymbol{h}$ is a local minimum point.

*B. Local mean stability*

The local stability of the TLMM-NSAF algorithm was analyzed in this section. For convenient analysis that converges to the vicinity of $\boldsymbol{h}$ after a sufficient number of iterations. Since the instantaneous value is used to replace the expected value (13), it will cause a gradient error, which is defined as:

$$\boldsymbol{r}(\boldsymbol{w}(z)) \triangleq \hat{\boldsymbol{g}}_{TLMM-NSAF}(\boldsymbol{w}(z)) - \boldsymbol{g}_{TLMM-NSAF}(\boldsymbol{w}(z)). \quad (23)$$

Substituting (23) into (14), the weight update can rewrite as

$$\boldsymbol{w}(z+1) = \boldsymbol{w}(z) - \mu \boldsymbol{g}_{TLMM-NSAF}(\boldsymbol{w}(z)) - \mu \boldsymbol{r}(\boldsymbol{w}(z)). \quad (24)$$

By subtracting $\boldsymbol{h}$ from both sides of equation (24), the update formula for the weight error vector is:

$$\tilde{\boldsymbol{w}}(z+1) = \tilde{\boldsymbol{w}}(z) + \mu \boldsymbol{g}_{TLMM-NSAF}(\boldsymbol{w}(z)) + \mu \boldsymbol{r}(\boldsymbol{w}(z)). \quad (25)$$

Where $\tilde{\boldsymbol{w}}(z) \overset{def}{=} \boldsymbol{h} - \boldsymbol{w}(z)$. Due to $J_{TLMM-NSAF}(\boldsymbol{w})$ is twice continuously differentiable near the straight line between $\boldsymbol{h}$ and $\boldsymbol{w}(z)$ [17], [18], so employing Theorem 1.2.1 in the literature [22] the gradient can be calculated as

$$
\begin{aligned}
&\boldsymbol{g}_{TLMM-NSAF}(\boldsymbol{w}(z)) \\
&= \boldsymbol{g}_{TLMM-NSAF}(\boldsymbol{h}) - \left[\int_0^1 \boldsymbol{H}_{TLMM-NSAF}(\boldsymbol{h} - t\tilde{\boldsymbol{w}}(z))dt\right]\tilde{\boldsymbol{w}}(z) \\
&\approx \left[\int_0^1 \boldsymbol{H}_{TLMM-NSAF}(\boldsymbol{h})dt\right]\tilde{\boldsymbol{w}}(z) \\
&= -\boldsymbol{H}_{TLMM-NSAF}(\boldsymbol{h})\tilde{\boldsymbol{w}}(z),
\end{aligned} \tag{26}
$$

since $\tilde{\boldsymbol{w}}(z)$ is very small near the expected solution $\boldsymbol{h}$, an approximation $\boldsymbol{h} - t\tilde{\boldsymbol{w}}(z) \approx \boldsymbol{h}$ can be obtained.

The gradient error vector at the steady state can be approximated as

$$
\boldsymbol{r}(\boldsymbol{w}(z)) \approx \boldsymbol{r}(\boldsymbol{h}) = \hat{\boldsymbol{g}}_{TLMM-NSAF}(\boldsymbol{h}). \tag{27}
$$

Substituting (26) and (27) into (25), can get

$$
\begin{aligned}
\tilde{\boldsymbol{w}}(z+1) &\approx \tilde{\boldsymbol{w}}(z) - \mu \boldsymbol{H}_{TLMM-NSAF}(\boldsymbol{h})\tilde{\boldsymbol{w}}(z) + \mu \boldsymbol{r}(\boldsymbol{h}) \\
&\approx (\boldsymbol{I} - \mu \boldsymbol{H}_{TLMM-NSAF}(\boldsymbol{h}))\tilde{\boldsymbol{w}}(z) + \mu \boldsymbol{r}(\boldsymbol{h}).
\end{aligned} \tag{28}
$$

Taking expectations on both sides of (28), we can obtain

$$
E[\tilde{\boldsymbol{w}}(z+1)] = (\boldsymbol{I} - \mu \boldsymbol{H}_{TLMM-NSAF}(\boldsymbol{h}))E[\tilde{\boldsymbol{w}}(z)]. \tag{29}
$$

Based on (29), in order to guarantee local convergence in the mean sense, the magnitude of all the eigenvalues of matrix $\boldsymbol{I} - \mu \boldsymbol{H}_{TLMM-NSAF}(\boldsymbol{h})$ is less than 1, So it can be expressed as

$$
\left|1 - \mu \lambda_{\max}\left\{\boldsymbol{H}_{TLMM-NSAF}(\boldsymbol{h})\right\}\right| < 1. \tag{30}
$$

where $\lambda_{\max}\{\boldsymbol{A}\}$ is defined as the largest eigenvalue of the matrix $\boldsymbol{A}$.

Substituting (20) into (30), the step size bound to ensure the TLS-NSAF algorithm stable convergence can be obtained as

$$
0 < \mu < \frac{2}{\lambda_{\max}\left\{\frac{1}{\|\boldsymbol{h}\|^2}\left[\sum_{i=0}^{N-1}\gamma_i \boldsymbol{I} - \sum_{i=0}^{N-1}\frac{\sigma_{i,in}^2 \boldsymbol{I}}{(L-2)\sigma_{i,\tilde{x}}^2}\right]\right\}}. \tag{31}
$$

However, in order to ensure the stable convergence of the TLMM-NSAF algorithm, the following analysis is required:

By subtracting $\boldsymbol{h}$ (14) from both sides of the equation, the update formula of the weight error vector can be rewritten as:

$$
\tilde{\boldsymbol{w}}(z+1) = \tilde{\boldsymbol{w}}(z) - \mu \hat{\boldsymbol{g}}_{TLMM-NSAF}(\boldsymbol{w}(z)) \tag{32}
$$

Take the squared Euclidean norm on both sides of (32). Rearranging terms, and taking the expectation, the update equation can be represented in terms of the mean-square deviation as

$$
\begin{aligned}
&MSD(z+1) \\
&= MSD(z) - 2\mu E\left[\tilde{\boldsymbol{w}}(z)^T \hat{\boldsymbol{g}}_{TLMM-NSAF}(\boldsymbol{w}(z))\right] \\
&+ \mu^2 E\left[\hat{\boldsymbol{g}}_{TLMM-NSAF}(\boldsymbol{w}(z))^T \hat{\boldsymbol{g}}_{TLMM-NSAF}(\boldsymbol{w}(z))\right]
\end{aligned} \tag{33}
$$

and $MSD(z)$ in (33) is defined as $MSD(z) \overset{def}{=} E\left[\|\tilde{\boldsymbol{w}}(z)\|^2\right]$. In order to ensure the stable convergence of the algorithm, it is necessary to satisfy

$$
MSD(z+1) - MSD(z) < 0 \tag{34}
$$

Substituting (34) into (33) yields the range of step size $\mu$ as:

$$
0 < \mu < \frac{2E\left[\tilde{\boldsymbol{w}}(z)^T \hat{\boldsymbol{g}}_{TLMM-NSAF}(\boldsymbol{w}(z))\right]}{E\left[\hat{\boldsymbol{g}}_{TLMM-NSAF}(\boldsymbol{w}(z))^T \hat{\boldsymbol{g}}_{TLMM-NSAF}(\boldsymbol{w}(z))\right]}. \tag{35}
$$

The paper [7] states that, consider the situation where the disturbance is negligible, the undisturbed error signal $\boldsymbol{h}^T \tilde{x}_i(z) - \boldsymbol{w}^T(z)\tilde{x}_i(z)$ is equal to the decimated subband error signal $e_{i,D}(z)$ and when the algorithm iterates enough, $\boldsymbol{w}(z) \approx \boldsymbol{h}$. So (35) can be simplified to

$$
0 < \mu < 2\left(\|\boldsymbol{h}\|^2 + \theta_i\right) \tag{36}
$$

The detailed derivation of (36) is provided in the Appendix A. Combining (32) and (33), the range of step sizes that make the algorithm stable can be obtained as:

$$
0 < \mu < \min\left(2\left(\|\boldsymbol{h}\|^2 + \theta_i\right), \frac{2}{\lambda_{\max}\left\{\boldsymbol{H}_{TLMM-NSAF}(\boldsymbol{h})\right\}}\right) \tag{37}
$$

## V. Steady-State Mean-Square Performance Analysis

In (13), the existence of gradient error B due to the use of instantaneous values instead of expected values (see Section IV above), the MSD of the algorithm converges to a nonzero steady state value. This section uses an energy conservation method to predict this steady-state MSD.

Taking the square of the weighted Euclidean norm on both sides of Equation (25), and calculating the expectation of each term, and get:

$$
\boldsymbol{E}\left[\|\tilde{\boldsymbol{w}}(z+1)\|_{\Lambda}^2\right] \approx \boldsymbol{E}\left[\|\tilde{\boldsymbol{w}}(z)\|_{\Theta}^2\right] + \mu^2 \boldsymbol{E}\left[\|\boldsymbol{r}(\boldsymbol{h})\|_{\Lambda}^2\right]. \tag{38}
$$

$\Lambda$ is a non-negative definite weighting matrix and $\Theta = \left[\boldsymbol{I} - \mu \boldsymbol{H}_{TLMM-NSAF}(\boldsymbol{w})\right]\Lambda\left[\boldsymbol{I} - \mu \boldsymbol{H}_{TLMM-NSAF}(\boldsymbol{w})\right]$. Furthermore, we define the square criterion of the weighted Euclidean norm of the vector $\boldsymbol{s}$ as $\|s\|_{\mathbf{A}}^2 \overset{def}{=} s^T \mathbf{A} s$, where $\mathbf{A}$ is the weighting matrix.

Next using the vectorization operation, two transformation properties between the trace of a matrix and the Kronecker product [28], which are defined as follows:

$$
\begin{aligned}
\boldsymbol{tr}\left\{\boldsymbol{X}^T \boldsymbol{Y}\right\} &= vec\left\{\boldsymbol{Y}\right\}^T vec\left\{\boldsymbol{X}\right\}, \\
vec\left\{\boldsymbol{XYZ}\right\} &= \left(\boldsymbol{Z}^T \otimes \boldsymbol{X}\right) vec\left\{\boldsymbol{Y}\right\}.
\end{aligned}
$$

Where $\boldsymbol{X}$, $\boldsymbol{Y}$, $\boldsymbol{Z}$ are all matrices, $\otimes$ is the Kronecker product and $vec\{.\}$ is the vectorization operator that stacks the columns of its matrix argument into a single column vector.

In fact, matrix $\Lambda$ is symmetric and deterministic. Applying the above two properties allows the following simplification operations:

$$
\begin{aligned}
\boldsymbol{E}\left[\|\boldsymbol{r}(\boldsymbol{h})\|_{\Lambda}^2\right] &= \boldsymbol{E}\left[\boldsymbol{tr}\left\{\Lambda \boldsymbol{r}(\boldsymbol{h})\boldsymbol{r}(\boldsymbol{h})^T\right\}\right] \\
&= \boldsymbol{tr}\left\{\Lambda E\left[\boldsymbol{r}(\boldsymbol{h})\boldsymbol{r}(\boldsymbol{h})^T\right]\right\} \\
&= \boldsymbol{tr}\left\{\Lambda \boldsymbol{M}\right\} \\
&= vec\left\{\boldsymbol{M}\right\}^T vec\left\{\Lambda\right\} \\
&\overset{def}{=} \boldsymbol{m}^T \boldsymbol{n}
\end{aligned} \tag{39}
$$

where $\boldsymbol{m} \overset{def}{=} vec\{\boldsymbol{M}\}$, $\boldsymbol{n} \overset{def}{=} vec\{\Lambda\}$,

$$\boldsymbol{M} = E\left[\boldsymbol{r}(\boldsymbol{h})\boldsymbol{r}(\boldsymbol{h})^T\right]$$
$$= \sum_{i=0}^{N-1} \frac{\sigma_{i,in}^2}{\|\bar{\boldsymbol{h}}\|} \frac{\gamma_i}{(L-4)\sigma_{i,\tilde{x}}^2} \boldsymbol{I} - \sum_{i=0}^{N-1} \frac{1}{\|\bar{\boldsymbol{h}}\|^4} \frac{3\sigma_{i,\tilde{x}}^4}{(L-2)(L-4)\sigma_{i,\tilde{x}}^4} \boldsymbol{h}\boldsymbol{h}^T \tag{40}$$

and

$$vec\{\Theta\}$$
$$= vec\left\{\left[\boldsymbol{I} - \mu \boldsymbol{H}_{TLMM-NSAF}(\boldsymbol{w})\right]\Lambda\left[\boldsymbol{I} - \mu \boldsymbol{H}_{TLMM-NSAF}(\boldsymbol{w})\right]\right\}$$
$$= \left(\left[\boldsymbol{I} - \mu \boldsymbol{H}_{TLMM-NSAF}(\boldsymbol{w})\right] \otimes \left[\boldsymbol{I} - \mu \boldsymbol{H}_{TLMM-NSAF}(\boldsymbol{w})\right]\right) vec\{\Lambda\}$$
$$\stackrel{def}{=} \boldsymbol{Pn}. \tag{41}$$

The detailed derivation of (40) is provided in the Appendix B. By substituting formulas (39) and (40) into formula (38), the following mean square relationship can be obtained:

$$\boldsymbol{E}\left[\|\tilde{\boldsymbol{w}}(z+1)\|_{\boldsymbol{n}}^2\right] \approx \boldsymbol{E}\left[\|\tilde{\boldsymbol{w}}(z)\|_{\boldsymbol{Pn}}^2\right] + \mu^2 \boldsymbol{m}^T \boldsymbol{n}. \tag{42}$$

According to the literature [1], [17], if the matrix $\boldsymbol{P}$ is stable, the equation (42) can be guaranteed to converge to the steady state. Therefore, when the step size $\mu$ of the selected algorithm satisfies the formula (31), the matrix $\boldsymbol{P}$ is stable, and the mean and mean square stability of the TLMM-NSAF algorithm can be satisfied.

When the algorithm iterates enough times, equation (42) can be rewritten as:

$$\boldsymbol{E}\left[\|\tilde{\boldsymbol{w}}(\infty)\|_{\boldsymbol{n}}^2\right] \approx \boldsymbol{E}\left[\|\tilde{\boldsymbol{w}}(\infty)\|_{\boldsymbol{Pn}}^2\right] + \mu^2 \boldsymbol{m}^T \boldsymbol{n}. \tag{43}$$

Then, we set $\boldsymbol{n}$ to $(\boldsymbol{I}-\boldsymbol{P})^{-1} vec\{\boldsymbol{I}\}$ and get the steady-state mean squared deviation of the TLMM-NSAF algorithm as:

$$\boldsymbol{E}\left[\|\tilde{\boldsymbol{w}}(\infty)\|^2\right] \approx \mu^2 \boldsymbol{m}^T (\boldsymbol{I}-\boldsymbol{P})^{-1} vec\{\boldsymbol{I}\}. \tag{44}$$

## VI. The Variable Step-Size Strategies

To address the problem that the fixed step size of adaptive filtering algorithms cannot balance fast convergence and low steady-state estimation, a convex combination scheme of VSS [29] [18] and two independently operating filters [30] has been proposed. This section combines the advantages of VSS with a convex combination scheme, and propose an improved variable stride strategy,The specific strategies are as follows:

$$\boldsymbol{w}(z) = \lambda(z)\boldsymbol{w}_1(z) + (1-\lambda(z))\boldsymbol{w}_2(z) \tag{45}$$

where $\boldsymbol{w}_1(z)$ and $\boldsymbol{w}_2(z)$ are two component filter weights, $\lambda(z)$ is a mixing parameter, and the overall filter weight $\boldsymbol{w}(z)$ is obtained by combining $\boldsymbol{w}_1(z)$ and $\boldsymbol{w}_2(z)$ through $\lambda(z)$. $\boldsymbol{w}_1(z)$ can be obtained by

$$\boldsymbol{w}_1(z+1) = \boldsymbol{w}_1(z) - \mu_{vss}(z)\hat{\boldsymbol{g}}_{TLMM-NSAF}(\boldsymbol{w}_1(z))$$
$$= \begin{cases} \boldsymbol{w}_1(z) + \mu_{vss}(z) \sum_{i=0}^{N-1} \frac{\|\bar{\boldsymbol{w}}_1(z)\|^2 e_{i,D}(z)\tilde{\boldsymbol{x}}_i(z) + e_{i,D}^2(z)\boldsymbol{w}_1(z)}{\|\bar{\boldsymbol{w}}_1(z)\|^4 (L-2)\sigma_{i,\tilde{x}}^2} & |e_{i,D}(z)| < \xi_i \\ \boldsymbol{w}_1(z) & |e_{i,D}(z)| \geq \xi_i \end{cases} \tag{46}$$

$\boldsymbol{w}_2(z)$ can be obtained by

$$\boldsymbol{w}_2(z+1) = \boldsymbol{w}_2(z) - \mu_2(z)\hat{\boldsymbol{g}}_{TLMM-NSAF}(\boldsymbol{w}_2(z))$$
$$= \begin{cases} \boldsymbol{w}_2(z) + \mu_2(z) \sum_{i=0}^{N-1} \frac{\|\bar{\boldsymbol{w}}_2(z)\|^2 e_{i,D}(z)\tilde{\boldsymbol{x}}_i(z) + e_{i,D}^2(z)\boldsymbol{w}_2(z)}{\|\bar{\boldsymbol{w}}_2(z)\|^4 (L-2)\sigma_{i,\tilde{x}}^2} & |e_{i,D}(z)| < \xi_i \\ \boldsymbol{w}_2(z) & |e_{i,D}(z)| \geq \xi_i \end{cases} \tag{47}$$

and

$$\mu_{vss}(z+1) = \begin{cases} \alpha\mu_{vss}(z) + \beta\sum_{i=0}^{N-1} e_{i,D}^2(z) & |e_{i,D}(z)| < \xi_i \\ \mu_{vss}(z) & |e_{i,D}(z)| \geq \xi_i \end{cases} \tag{48}$$

where $0<\alpha<1, \beta>0$ , $\mu_2(z)$ is a fixed small constant. Considering the stability of the algorithm and the required tracking capability, we need to set a range for the step size $\mu_{vss}(z)$:

$$\mu_{vss}(z+1) = \begin{cases} \mu_{\max}, & \mu(z+1) > \mu_{\max} \\ \mu_{vss}(z+1), & \textbf{\textit{otherwise}} \\ \mu_{\min}, & \mu(z+1) < \mu_{\min} \end{cases} \tag{49}$$

where $\mu_{\max}$ and $\mu_{\min}$ are the upper and lower bounds of the variable step size, respectively.The subband output signals of the overall filter are expressed as

$$y_{i,D}(z) = \lambda(z)y_{i,D,1}(z) + (1-\lambda(z))y_{i,D,2}(z) \text{ for } i \; [0, N-1]$$

where $y_{i,D,1}(z)$ and $y_{i,D,2}(z)$ denote the subband output signals of both component filters, respectively, which are given by

$$y_{i,D,1}(z) = \tilde{\boldsymbol{x}}_i^T(z)\boldsymbol{w}_1(z) \text{ and } y_{i,D,2}(z) = \tilde{\boldsymbol{x}}_i^T(z)\boldsymbol{w}_2(z).$$

Then, the overall subband error signals can be given by

$$e_{i,D}(z) = d_{i,D}(z) - y_{i,D}(z). \tag{50}$$

Evidently, the most critical problem of convex combination is how to adjust the mixing parameter $\lambda(z)$, which can be obtained by the following formula:

$$\lambda(z) = \frac{1}{1+\mathrm{e}^{-\alpha(z)}}. \tag{51}$$

where the auxiliary variable $\alpha(z)$ is updated by minimizing the cost function

$$J(z) = \sum_{i=0}^{N-1} e_{i,D}^2(z) \tag{52}$$

Taking the gradient of $J(z)$ with respect to $\alpha(z)$ gives:

$$\nabla J(z) = \frac{\partial J(z)}{\partial \alpha(z)}$$
$$= \lambda(z)\left[1-\lambda(z)\right]\sum_{i=0}^{N-1} e_{i,D}(z)\left(y_{i,D,1}(z) - y_{i,D,2}(z)\right) \tag{53}$$

Next, we adapt $\alpha(z)$ using the normalized gradient method, obtaining

$$\alpha(z+1) = \alpha(z) - \mu_a \frac{\nabla J(z)}{\|\nabla J(z)\|}$$
$$= \alpha(z) + \mu_a \operatorname{sgn}\left[\sum_{i=0}^{N-1} e_{i,D}(z)\left(y_{i,D,1}(z) - y_{i,D,2}(z)\right)\right] \tag{54}$$

where $\mu_a$ is a step size of $\alpha(z)$ and $\mathrm{sgn}[.]$ is the sign function. Usually $\alpha(z)$ is restricted to the $[-a,a]$ range and the value of $a$ is set to 4 [31].

In this way, the improved VSS-CTLMM-NSAF algorithm can not only maintain a fast convergence speed in the whole process of algorithm convergence, but also further reduce the steady-state error. A simulation comparison of the Variable step strategies total least squares normalized subband adaptive filter (VSS-TLMM-NSAF) algorithm, the Convex combination total least squares normalized subband adaptive filter (CTLMM-NSAF) algorithm and the VSS-CTLMM-NSAF algorithm is performed in Section VIII. Simulation results demonstrate the superiority of our proposed variable step size strategy. The pseudocode of the proposed VSS-CTLMM-NSAF algorithm is summarized in TABLE II.

TABLE II
THE PSEUDOCODE OF THE VSS-CTLMM-NSAF ALGORITHM.

**Initialization:** $w_1(0)=w_2(0)=0$, $a^+=4$, $a(0)=0$

**Parameters:** $k=1.483\left(1+5/\left(N_w-1\right)\right)$, $j=1,2$, $\mu$, $N_w$, $\lambda_\sigma$

**For** z= 0,1,2, ...

**For** i=[0,N-1]

$$y_{j,i,D}(z)=\tilde{\boldsymbol{x}}_i^T(z)\boldsymbol{w}_j(z)$$

$$e_{j,i,D}(z)=\tilde{d}_{i,D}(z)-\tilde{\boldsymbol{x}}_i^T(z)\boldsymbol{w}_j(z)$$

$$y_{i,D}(z)=\lambda(z)y_{1,i,D}(z)+(1-\lambda(z))y_{2,i,D}(z)$$

$$e_{i,D}(z)=\lambda(z)e_{1,i,D}(z)+(1-\lambda(z))e_{2,i,D}(z)$$

$$A_{j,e,i}(z)=\left[e_{j,i,D}^2(z),e_{j,i,D}^2(z-1),\dots,e_{j,i,D}^2(z-N_w+1)\right]$$

$$\hat{\sigma}_{j,e,i}^2(z)=\lambda_\sigma\hat{\sigma}_{j,e,i}^2(z-1)+k(1-\lambda_\sigma)med\left(A_{j,e,i}(z)\right)$$

$$\xi_{j,i}=2.576\hat{\sigma}_{j,e,i}(n)$$

**If** $\left|e_{1,i,D}(z)\right|<\xi_{1,i}$

$$\boldsymbol{w}_1(z+1)=\boldsymbol{w}_1(z)+\mu_{vss}\sum_{i=0}^{N-1}\frac{\left\|\overline{\boldsymbol{w}}_1(z)\right\|^2 e_{1,i,D}(z)\tilde{\boldsymbol{x}}_i(z)+e_{1,i,D}^2(z)\boldsymbol{w}_1(z)}{\left\|\overline{\boldsymbol{w}}_1(z)\right\|^4(L-2)\sigma_{i,\tilde{x}}^2}$$

$$\mu_{vss}(z+1)=\alpha\mu_{vss}(z)+\beta\sum_{i=0}^{N-1}e_{i,D}^2(z)$$

**else**

$$\boldsymbol{w}_1(z+1)=\boldsymbol{w}_1(z)$$

$$\mu_{vss}(z+1)=\mu_{vss}(z)$$

**End**

**If** $\left|e_{2,i,D}(z)\right|<\xi_{2,i}$

$$\boldsymbol{w}_2(z+1)=\boldsymbol{w}_2(z)+\mu_2\sum_{i=0}^{N-1}\frac{\left\|\overline{\boldsymbol{w}}_2(z)\right\|^2 e_{2,i,D}(z)\tilde{\boldsymbol{x}}_i(z)+e_{2,i,D}^2(z)\boldsymbol{w}_2(z)}{\left\|\overline{\boldsymbol{w}}_2(z)\right\|^4(L-2)\sigma_{i,\tilde{x}}^2}$$

**else**

$$\boldsymbol{w}_2(z+1)=\boldsymbol{w}_2(z)$$

**End**

**End**

**Proposed combination scheme**

$$\alpha(z+1)=\alpha(z)-\mu_a\,\mathrm{sgn}\{\sum_{i=0}^{N-1}e_{i,D}(z)[y_{2,i,D}(z)-y_{1,i,D}(z)]\}$$

$$\alpha(z+1)=\min[\max(\alpha(z+1),-\alpha^+),\alpha^+]$$

$$\lambda(z+1)=\frac{1}{1+e^{-\alpha(z+1)}}$$

$$\boldsymbol{w}(z+1)=\lambda(z+1)\boldsymbol{w}_1(z+1)+(1-\lambda(z+1))\boldsymbol{w}_2(z+1)$$

**End**

## VII. COMPLEXITY ANALYSIS

The computational complexity of the NLMS, NSAF, M-NSAF, TLS-NSAF TLMM-NSAF VSS-TLMM-NSAF CTLMM-NASF and VSS-CTLMM-NSAF algorithms are presented in Table III during each iteration. In this table, $M$ expresses the length of the analysis filter; $N$ means the number of subband filter; $L$ denotes the length of the subband adaptive filter. As can be seen from Table III, the computational complexity of the TLMM-NSAF algorithm is increased compared to the TLS-NSAF algorithm. But the increased computational complexity is relatively small and acceptable in practical applications. On the other hand, the increased computational complexity is also worthwhile, as the proposed algorithm significantly enhances the system's resistance to impulse noise disturbances. Compared with the fixed-step algorithm, the computational complexity of the traditional convex combination and the variable-step algorithm has increased to a certain extent. Compared with the traditional convex combination algorithm, the computational complexity of the improved variable-step algorithm is also slightly increased. But the increased computational complexity is worth it, because the performance of the improved algorithm has been significantly improved.

TABLE III
COMPUTATIONAL COMPLEXITY

| Algorithm | Multiplications | Additions | Divisions | Squearoot | Exponent |
|---|---|---|---|---|---|
| NLMS | $3L+1$ | $3L-1$ | 1 | 0 | 0 |
| NSAF | $3L+NM$ $+2M+2$ | $3L+N(M-1)+$ $2(M-1)$ | 1 | 0 | 0 |
| M-NSAF | $5L+NM+$ $2M+7$ | $5L+N(M-1)+$ $2(M-1)+3$ | 2 | 0 | 0 |
| TLS-NSAF | $6L+NM+$ $2M+3$ | $5L+N(M-1)+$ $2(M-1)$ | 3 | 0 | 0 |
| TLMM-NSAF | $6L+N(N_w+5)$ $+NM+2M+3$ | $5L+N(M+3)+$ $2(M-1)$ | 4 | 1 | 0 |
| VSS-TLMM-NSAF | $6L+N(N_w+8)$ $+NM+2M+3$ | $5L+N(M+4)+$ $2(M-1)$ | 4 | 1 | 0 |
| CTLMM-NSAF | $12L+2N(N_w+5)$ $+4NM+4M+11$ | $10L+2N(M+3)+$ $4M+(M+4)/N$ | $4+\frac{1}{N}$ | 1 | $\frac{1}{N}$ |
| VSS-CTLMM-NSAF | $12L+2N(N_w+8)$ $+4NM+4M+11$ | $10L+2N(M+4)+$ $4M+(M+4)/N$ | $4+\frac{1}{N}$ | 1 | $\frac{1}{N}$ |

## VIII. SIMULATIONS

This section verifies the previously proposed algorithm and theoretical results through simulations. All curves were obtained by averaging the results of 100 independent experiments, unless otherwise stated.

### A. *Number of different subbands*

This subsection compares the performance of the TLMM-NSAF algorithm with different numbers of subbands. Generate an unknown weight vector $\boldsymbol{h}$ by using a uniform distribution from -0.5 to 0.5,and $L=512(\|\boldsymbol{h}\|^2=1)$ .The length of the subband adaptive filter adopts the same length $(L=512)$ as the unknown weight vector $\boldsymbol{h}$. The input signal is the correlated input signal, which uses an Autoregressive(AR) model and it can get by filtering a zero-mean Gaussian signal through $T_1(c)=1/(1-0.8c^{-1})$ [16].

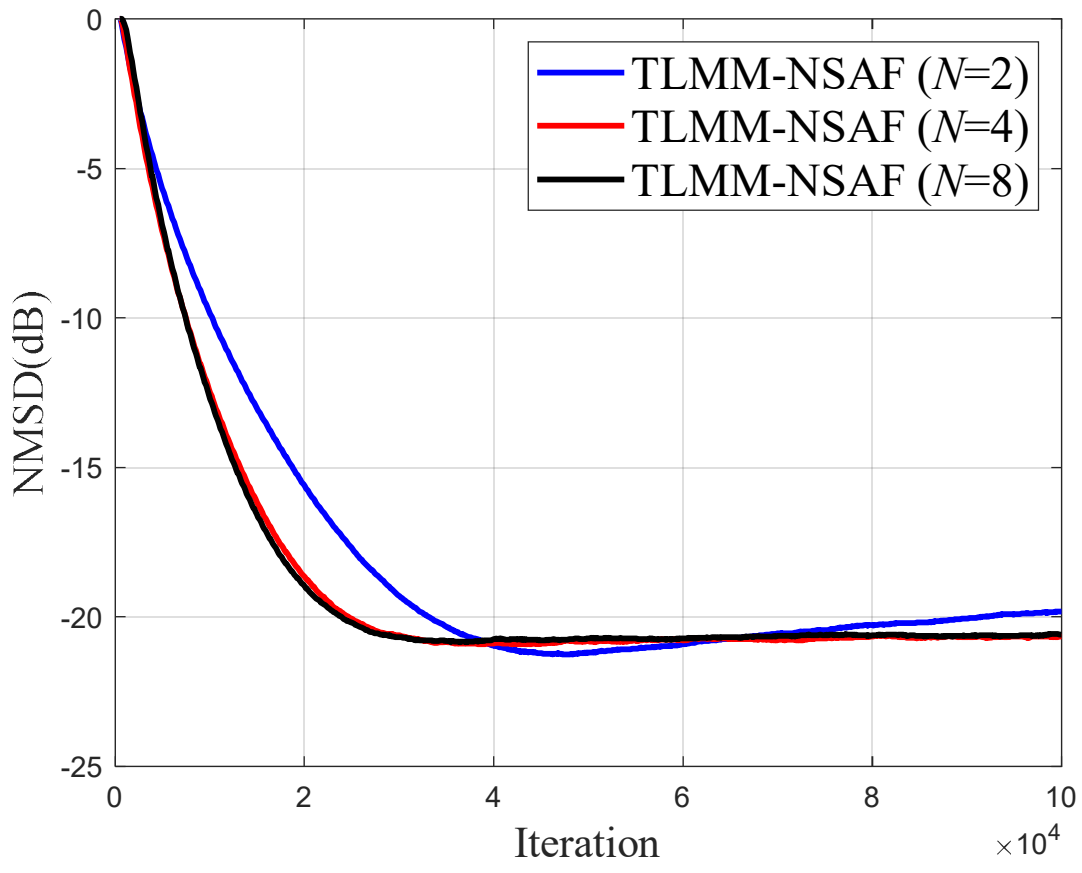

Fig. 2. TLMM-NSAF algorithm curves under different numbers of subband, without impulse noise

( $\sigma_{in}^2=0.05,\sigma_o^2=0.05,\mu=0.13$ )

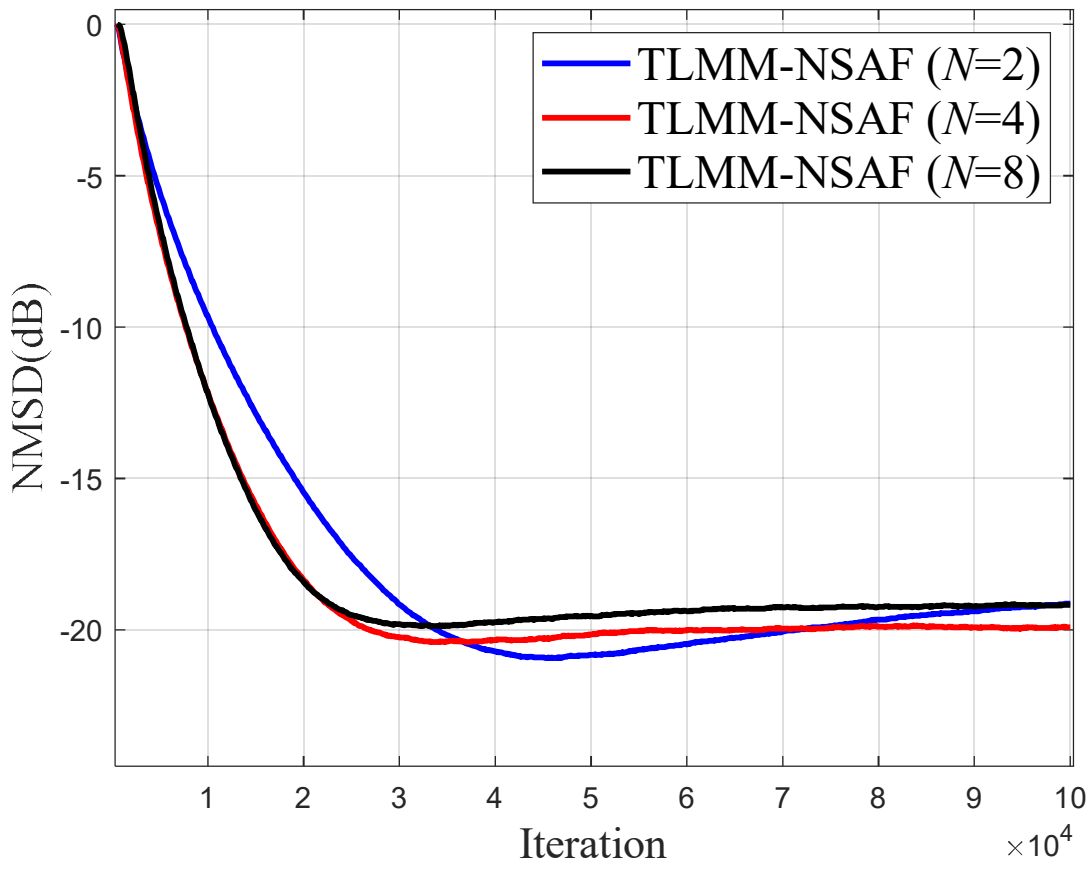

Fig. 3. TLMM-NSAF algorithm curves under different numbers of subband, contains impulse noise

( $\sigma_{in}^2=0.05,\sigma_o^2=0.05,\mu=0.13$ )

For ease of comparison, we choose an appropriate step size parameter $\mu$ to ensure that the algorithm has the same convergence rate in early iterations. The algorithm performance evaluation index selects normalized mean square deviation (NMSD), which is defined as $NMSD=10\log_{10}\left(\|\boldsymbol{w}(z)-\boldsymbol{h}\|^2/\|\boldsymbol{h}\|^2\right)$.

From Figure 2 and Figure 3, it can concluded that the more the number of subbands, the faster the algorithm convergence rate, but the NMSD will increase. When the number of subbands exceeds 4, the convergence speed of the algorithm does not increase significantly, and the computational complexity also increases. Therefore， in order to obtain suitable algorithm convergence speed and steady-state mean square error, the number of subbands 4 is selected.

### B. *Step size range*

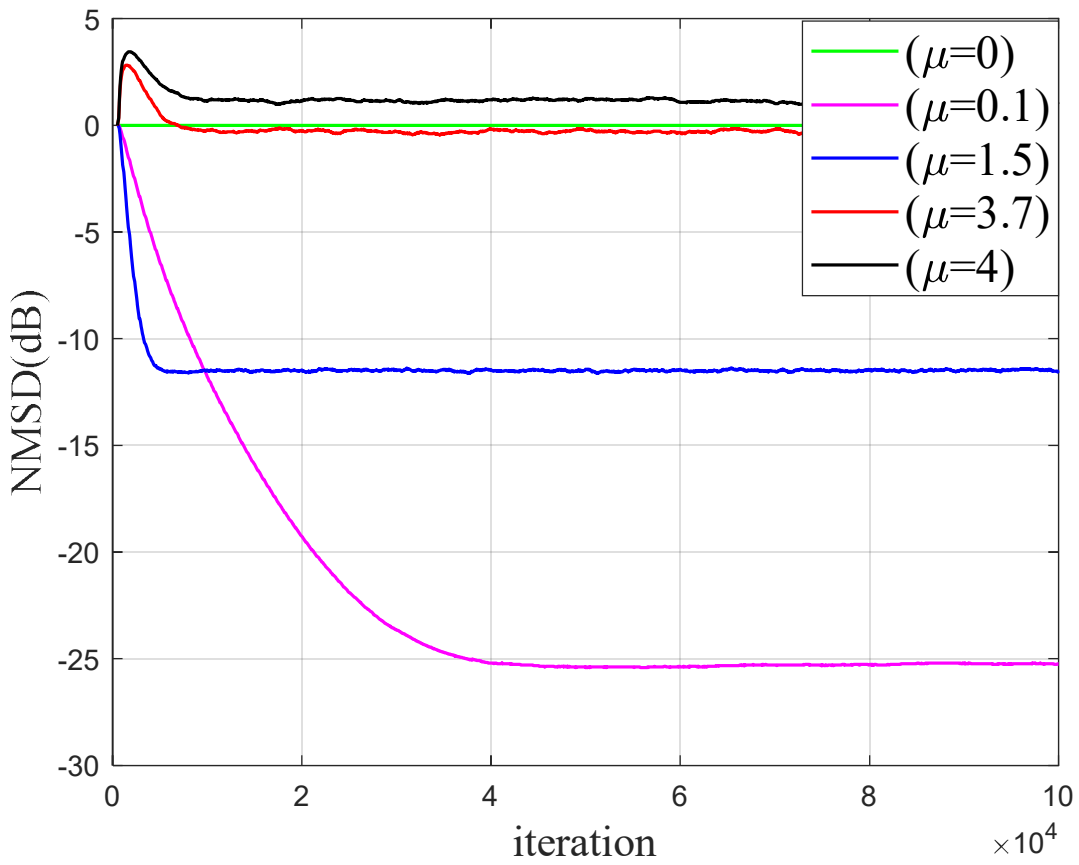

Fig. 4. TLMM-NSAF algorithm curve under different step lengths

( $\sigma_{in}^2=0.05,\sigma_o^2=0.05,N=4$ )

From Figure 4, due to $\sigma_{in}^2=0.05,\sigma_o^2=0.05$ , we can calculate the theoretical step size range of [0, 4] to ensure the stability of the algorithm according to formula (37). It is worth mentioning that because of the use of some acceptable approximations, the range of step sizes that actually guarantees the stability of the algorithm will be slightly less than 4. The simulation results confirm our conclusion.

### C. *Comparison results of different algorithms under relevant inputs*

It can be seen from Figure 5 and Figure 6 that the NLMS algorithm and the NSAF algorithm have poor convergence in the case of impulse noise, and the performance of the TLS-NSAF algorithm will be seriously deteriorated. The performance of M-NSAF algorithm and TLMM-NSAF algorithm is not disturbed by impulse noise. It can be seen from the simulation results that the performance of the proposed TLMM-NSAF algorithm is better than that of the M-NSAF algorithm. Furthermore, the performance of the VSS-CTLMM-NASF algorithm is optimal with or without impulse noise.

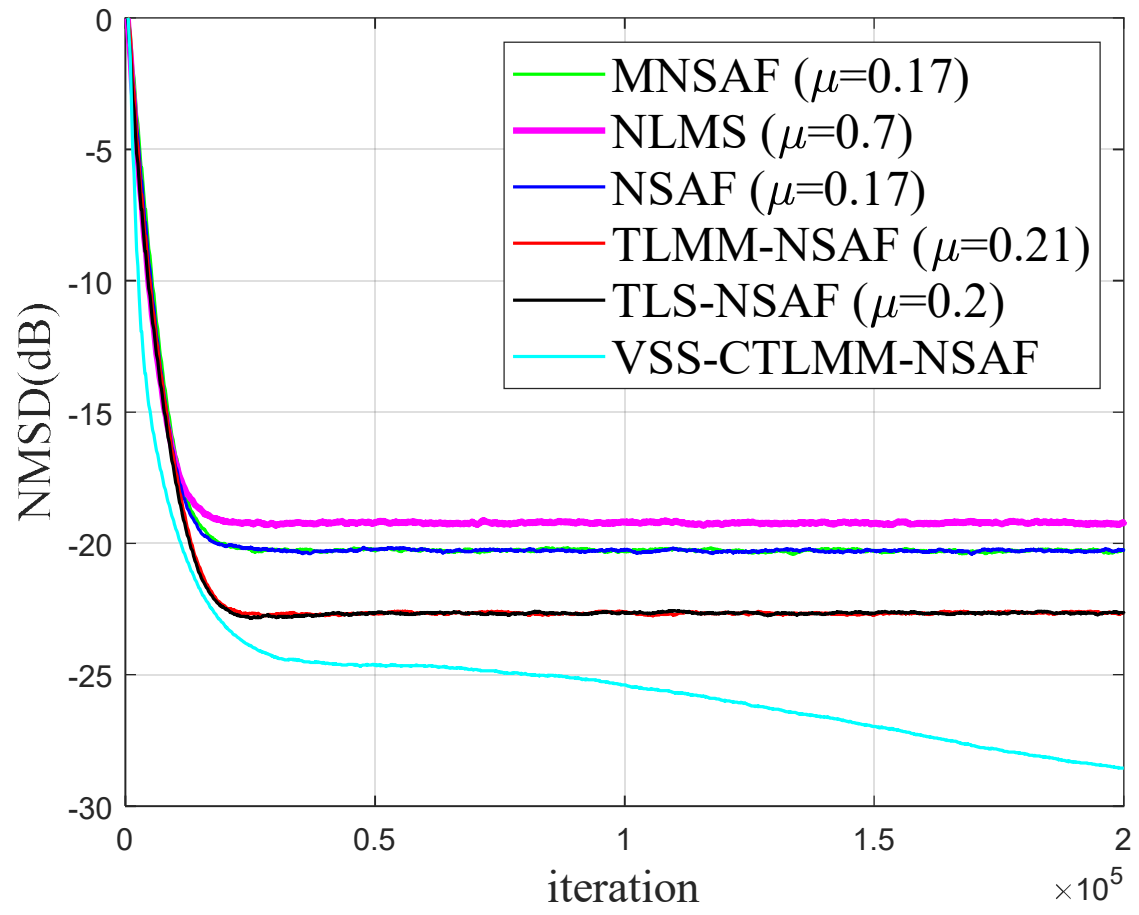

Fig. 5. Comparison curves of different algorithms under Gaussian noise $\sigma_{in}^2 = 0.05, \sigma_o^2 = 0.05, \theta_i = 1$ .

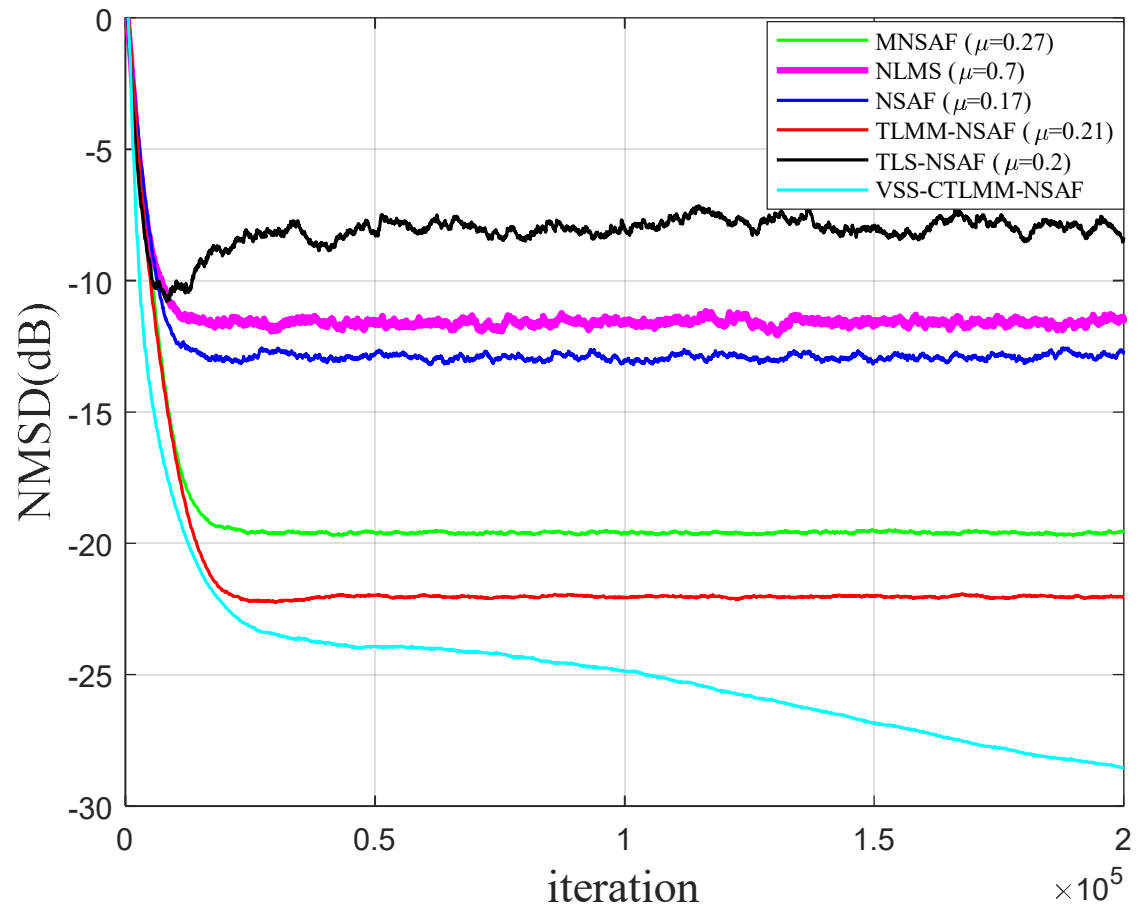

Fig. 6. Comparison curves of different algorithms under impulse noise

$\sigma_{in}^2 = 0.05, \sigma_o^2 = 0.05, \theta_i = 1$ .

### D. *Comparison of variable step size algorithms under different strategies*

As can be seen from Figure 7, the steady-state error of the convex

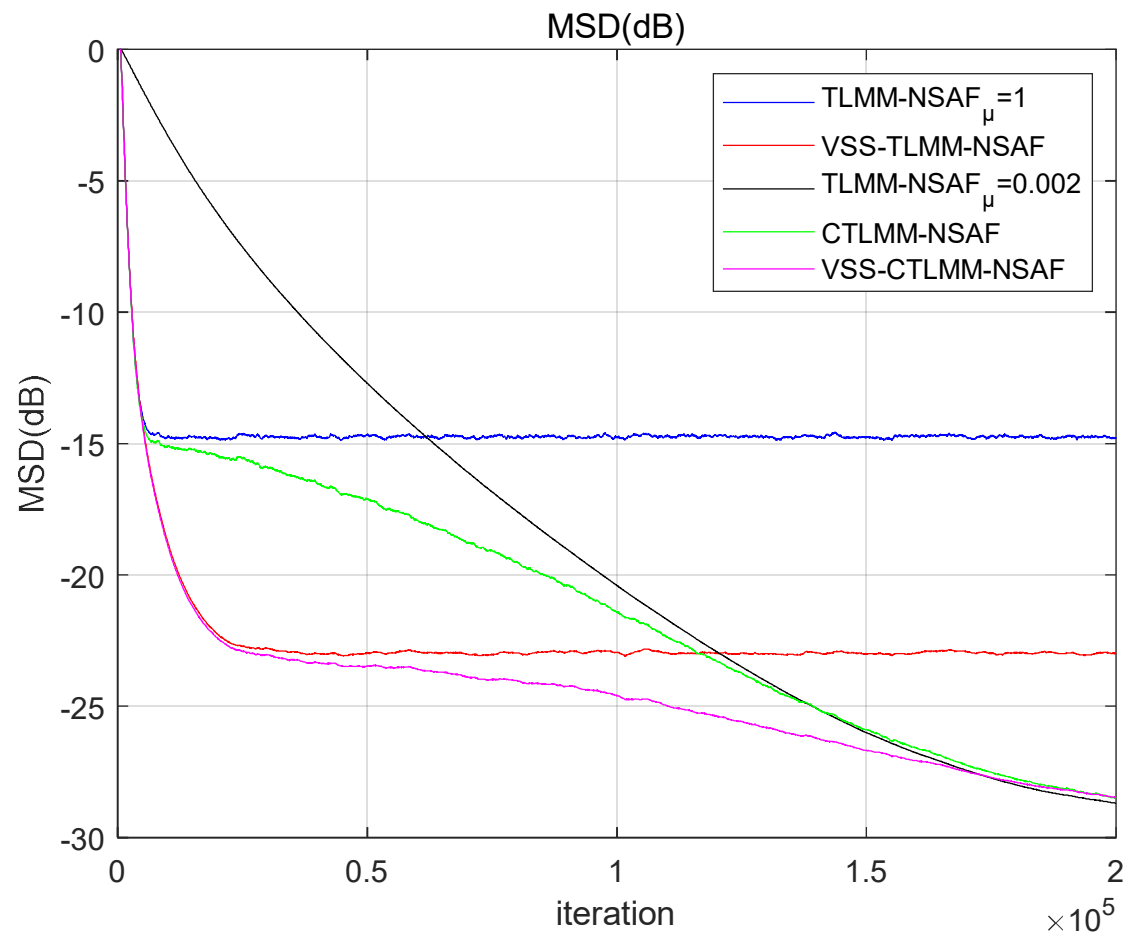

Fig. 7. Comparison curves of different algorithms under impulse noise

$\sigma_{in}^2 = 0.05, \sigma_o^2 = 0.05, \theta_i = 1, \alpha = 0.99, \beta = 0.0058$ .

combination strategy is better than the traditional variable-step algorithm, but the convergence speed is slower. The traditional variable-step algorithm has a faster convergence speed, but the steady-state error is larger than that of the small-step algorithm. Our proposed variable-step algorithm combines the advantages of convex combinations and traditional variable-step strategies, resulting in fast convergence and small steady-state errors. In contrast, the overall performance is better than other algorithms.

### E. *Algorithm theoretical performance verification*

The theoretical performance of the TLMM-NSAF algorithm is verified in this subsection. Due to the limitation of computer simulation software performance, the dimension of the unknown system weight vector used is set to order 128.

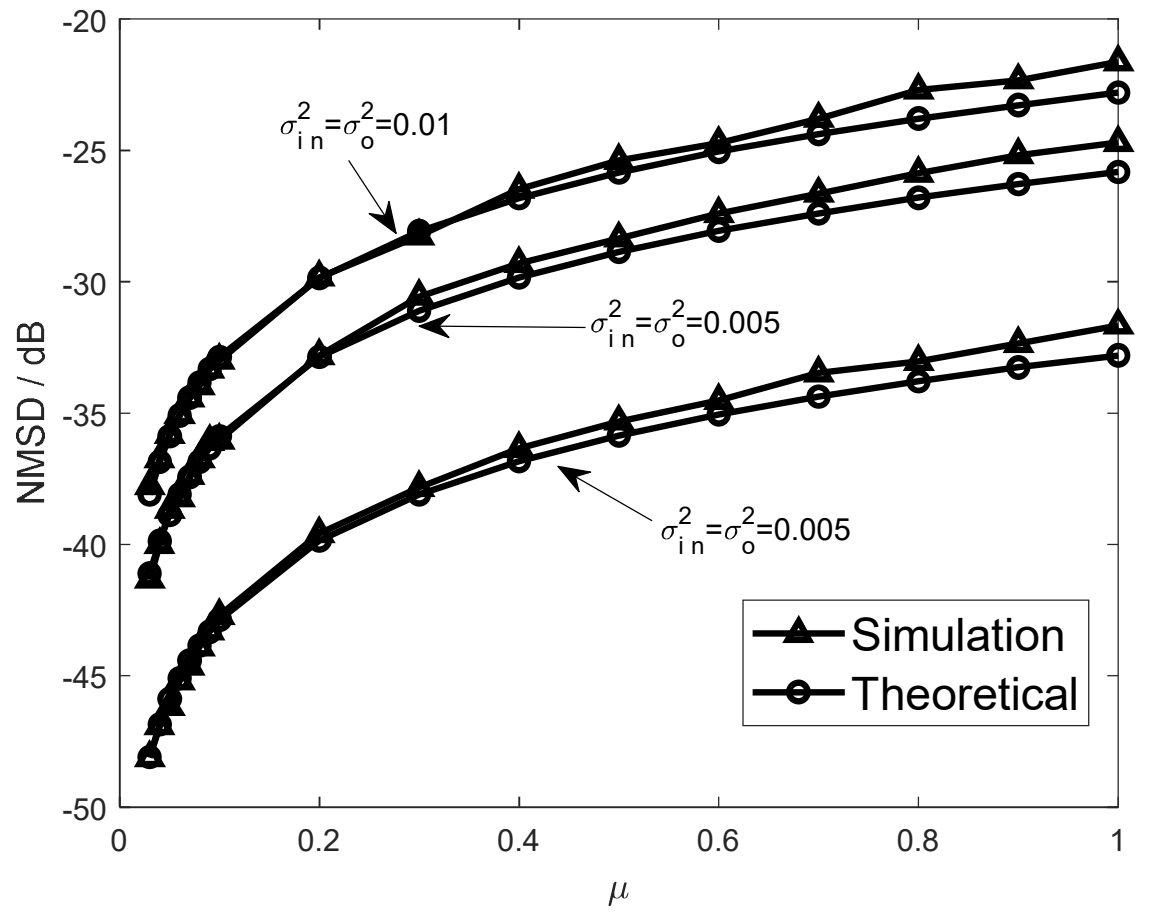

Fig. 8. Theoretical and simulated NMSD curves for different step lengths  and noise variances (Gaussian input)

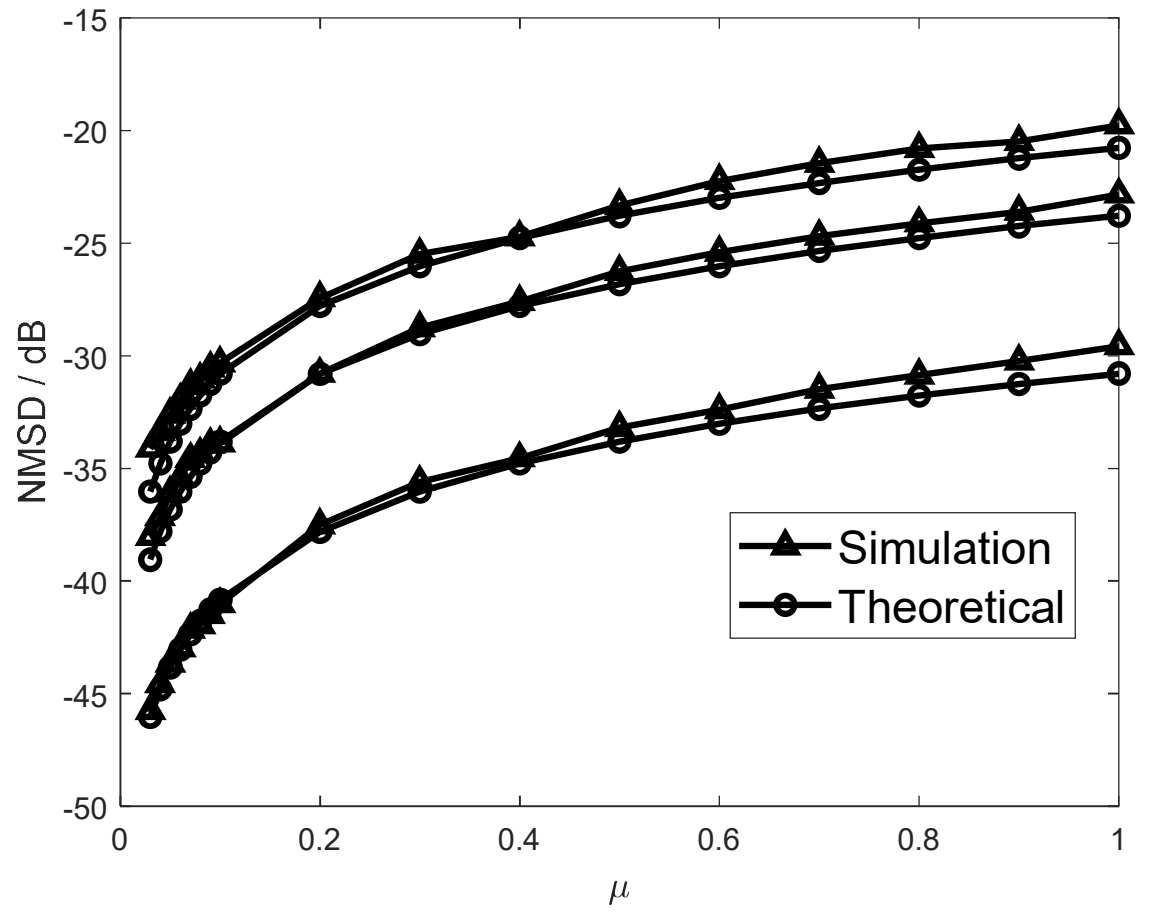

Fig. 9. Theoretical and simulated NMSD curves for different step lengths  and noise variances (Related input)

The simulation results in the experiments are the average after 10 independent runs. The theoretical steady state value is calculated from Equation (38).

Figure 8 and Figure 9 are the simulation curves under Gaussian input and correlation input, respectively. It can be seen from the simulation results that the theoretical curve of the proposed algorithm is basically consistent with the simulation curve under different variances. Although there are some minor differences, this is due to some assumptions and approximations made during the analysis, which are reasonable and acceptable. The comparison between the theoretical value curve and the simulation curve proves the accuracy of our analysis of the theoretical performance of the TLMM-NSAF algorithm.

F. *Acoustic echo cancellation*

This subsection verifies the convergence performance of the TLMM-NSAF algorithm in echo cancellation applications under real speech input.

Acoustic echo cancellation is commonly used in communication systems such as calls and video conferencing. Figure 10 shows the echo cancellation schematic.

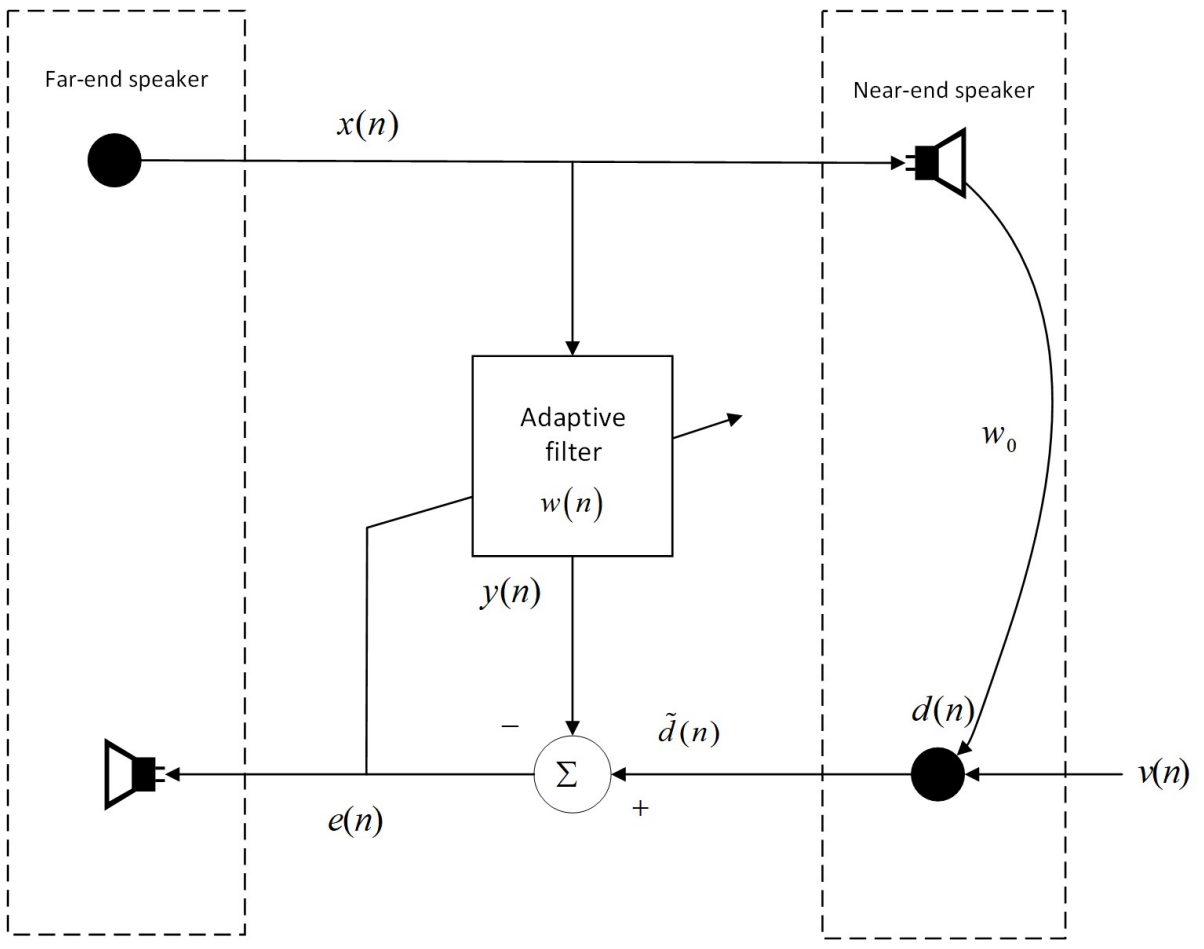

Fig. 10. Acoustic echo cancellation schematic

The application principle is as follows: $x(n)$ is the voice signal output by the far-end speaker, and also the input signal of the near-end speaker, $w_0$ is the pulse response of the echo channel, $d(n)$ is the echo signal generated by the input signal through the echo channel, and $v$ is the background noise. According to the adaptive filtering and estimation algorithm, the weight parameter $w(n)$ is continuously adjusted, and the echo path channel $w_0$ is gradually fitted. Finally, the echo signal value $y(n)$ estimated by the adaptive filter is close to the real echo signal. The estimated echo signal is subtracted from the near-end speech signal, and the obtained output error signal $e(n)=\tilde{d}(n)-y(n)$ will have no echo, and then be sent to the far-end, so as to achieve the purpose of echo cancellation.

Figure 11 is the voice input signal, Figure 12 is the echo signal, Figure 13 is the simulation results of different algorithms under the input of speech signal without impulse noise, and Figure 14 is the simulation result of different algorithms under the input of speech signal with impulse noise. It can be seen from the simulation results that the performance of the proposed TLMM-NSAF algorithm is comparable to that of the TLMM-NSAF algorithm without the interference of impulse noise, and is superior to other algorithms. In the case of including impulse noise, the performance of other algorithms The deterioration is serious, while the TLMM-NSAF algorithm shows excellent robustness. In addition, the simulation results show that the performance of the VSS-CTLMM-NSAF algorithm using the variable step size strategy has a certain improvement in the convergence speed and convergence compared with the TLMM-NSAF algorithm.

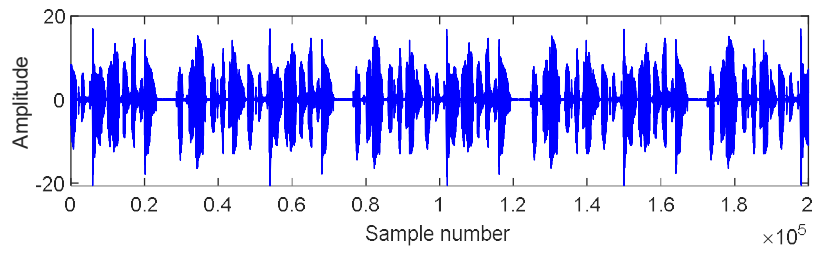

Fig. 11. Speech input signal

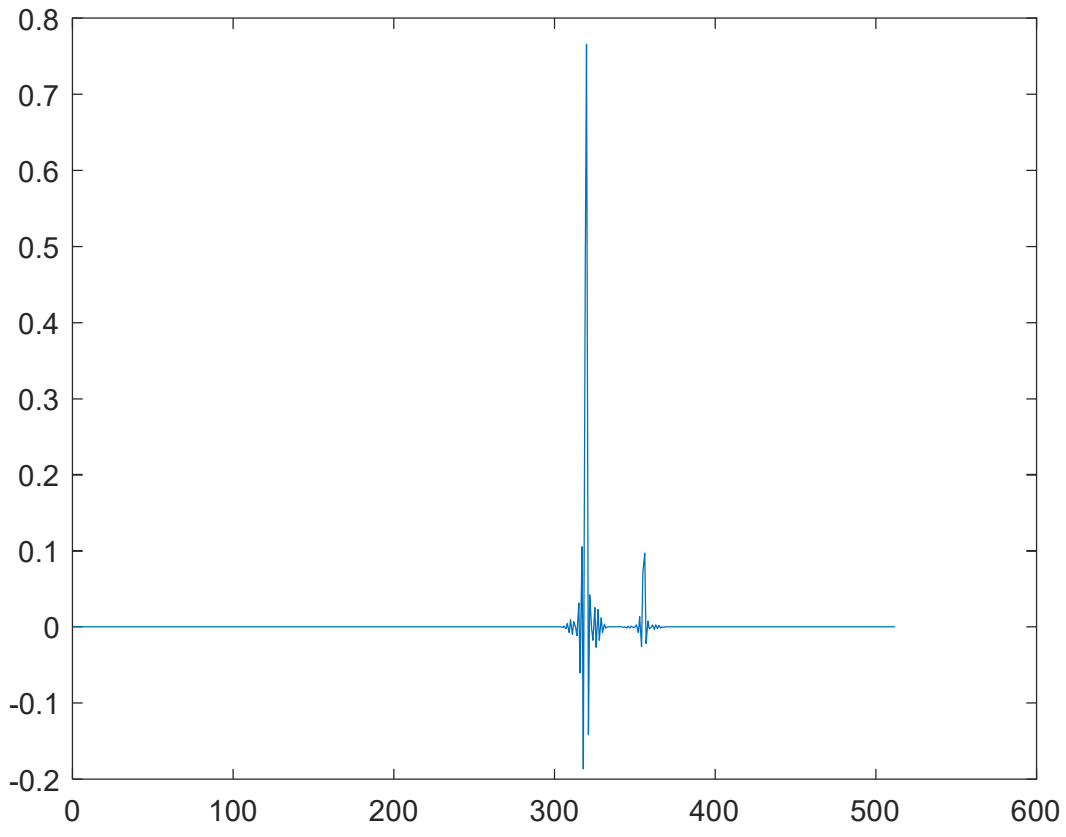

Fig. 12. Echo channel

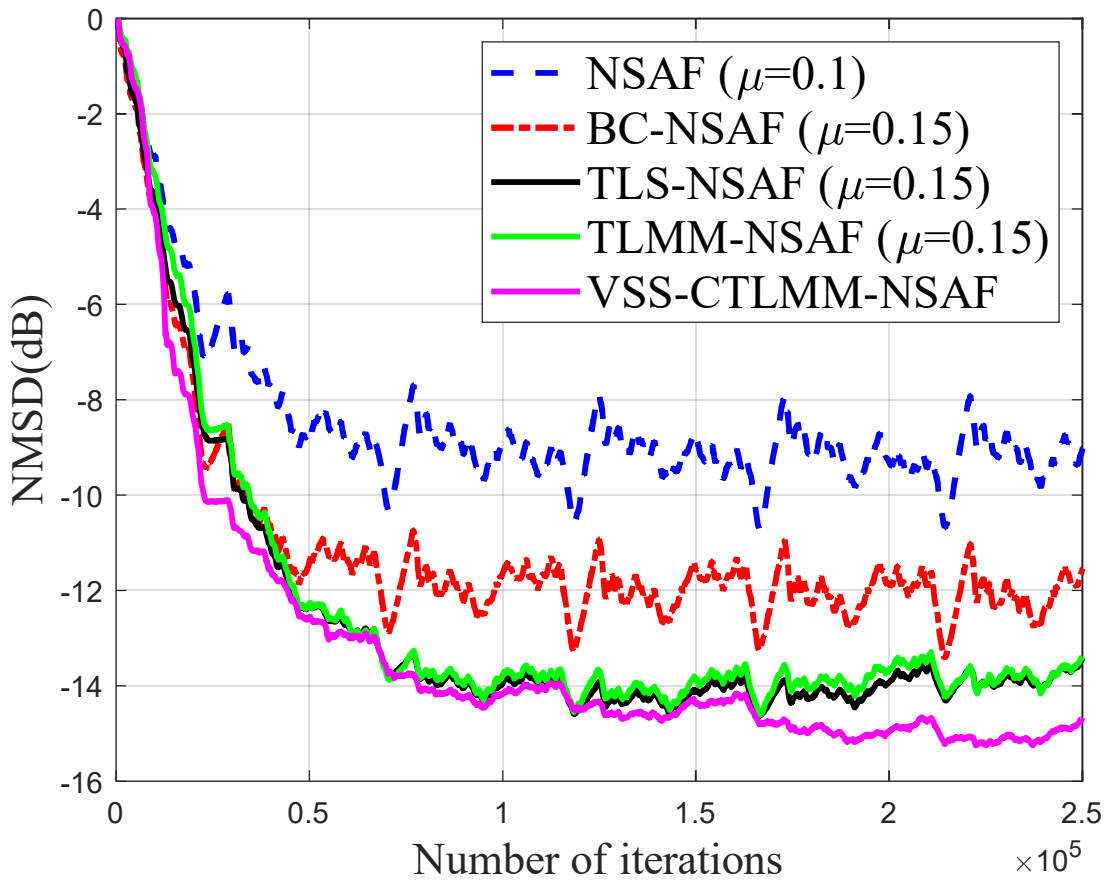

Fig. 13. Comparison of Steady-State Mean Square Deviations of different algorithms under speech input signal ( $\sigma_{in}^2=0.05, \sigma_o^2=0.05, \theta_i=1, Gaussian\ noise$ )

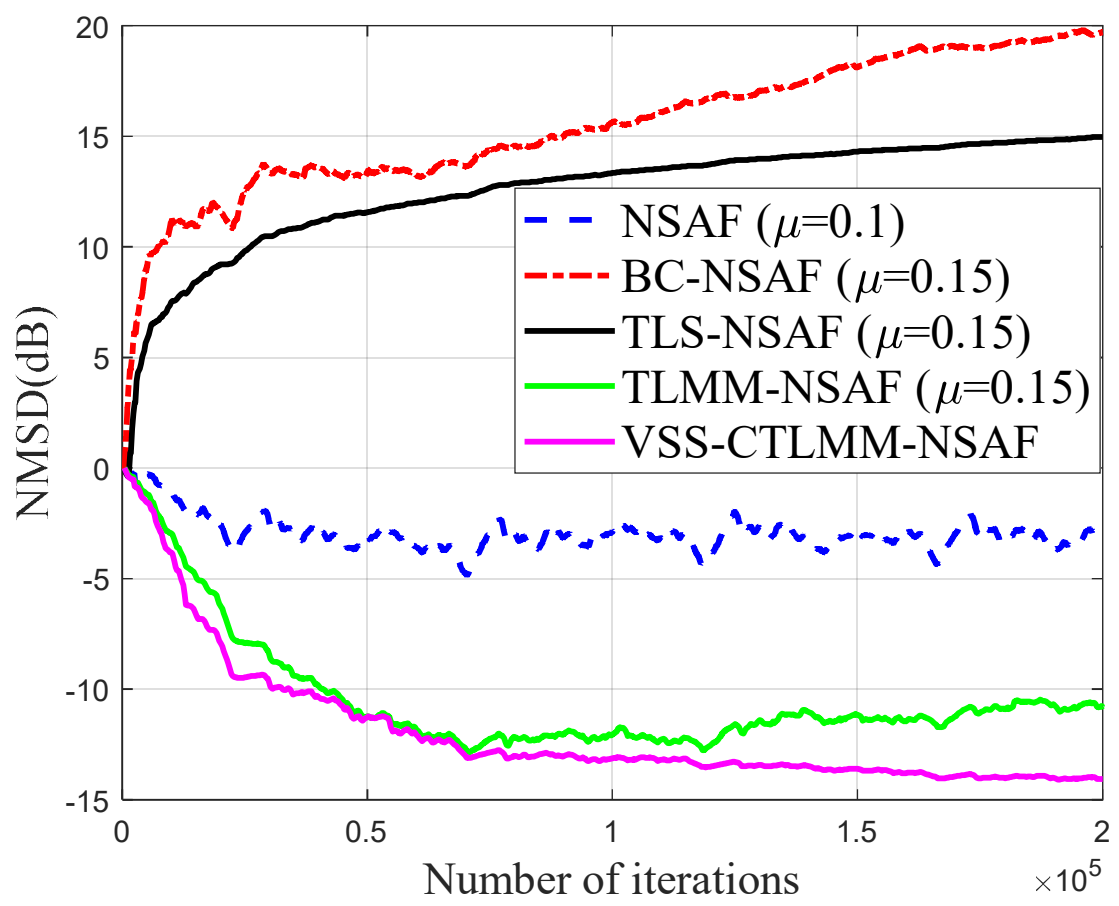

Fig. 14. Comparison of Steady-State Mean Square Deviations of different algorithms under speech input signal ( $\sigma_{in}^2=0.05, \sigma_o^2=0.05, \theta_i=1$, *Contains impulse noise* )

## IX. CONCLUSION

In this work, the proposed TLMM-NSAF algorithm addresses the severe performance degradation of traditional adaptive filtering algorithms under the EIV model disturbed by impulse noise. The detailed performance analysis of the TLMM-NSAF algorithm is carried out, and its analysis method also provides a good reference for the performance analysis of other sub-band adaptive filtering algorithms applying the TLS method in the future. In addition, in order to further improve the convergence speed and steady-state performance of the TLMM-NSAF algorithm, an improved VSS method is developed. Finally, we verify the convergence performance of the proposed TLMM-NSAF algorithm in system recognition and echo cancellation applications under real speech input, and demonstrate the superiority of our proposed algorithm by comparison with other adaptive filtering algorithms.

## APPENDIX A

Section B of Chapter IV of this paper has been given: $\boldsymbol{h}^T\tilde{x}_i(z)-\boldsymbol{w}^T(z)\tilde{x}_i(z)=e_{i,D}(z)$ and when the algorithm iterates enough, $\boldsymbol{w}(z)\approx\boldsymbol{h}$ . From this, (35) can be simplified to:

$$\frac{2E\left[\tilde{\boldsymbol{w}}(z)^T\hat{\boldsymbol{g}}_{TLMM-NSAF}(\boldsymbol{w}(z))\right]}{E\left[\hat{\boldsymbol{g}}_{TLMM-NSAF}(\boldsymbol{w}(z))^T\hat{\boldsymbol{g}}_{TLMM-NSAF}(\boldsymbol{w}(z))\right]} \approx \frac{2E\left[\dfrac{e_{i,D}^2}{\|\tilde{\boldsymbol{h}}\|^2\|\tilde{x}\|^2}\right]}{E\left[\dfrac{e_{i,D}^2\|\tilde{x}\|^2}{\|\tilde{\boldsymbol{h}}\|^4\|\tilde{x}\|^4}+\dfrac{2e_{i,D}^3\tilde{x}^T\boldsymbol{h}}{\|\tilde{\boldsymbol{h}}\|^6\|\tilde{x}\|^4}+\dfrac{e_{i,D}^4\boldsymbol{h}^T\boldsymbol{h}}{\|\tilde{\boldsymbol{h}}\|^8\|\tilde{x}\|^4}\right]} \tag{A1}$$

From the paper [13] we have $\|\tilde{x}\|^2\approx(L-2)\sigma_{i,\tilde{x}}^2$ and the length $L$ of the subband adaptive filter is large, so (A1) can be further simplified as:

$$\frac{2E\left[\tilde{\boldsymbol{w}}(z)^T\hat{\boldsymbol{g}}_{TLMM-NSAF}(\boldsymbol{w}(z))\right]}{E\left[\hat{\boldsymbol{g}}_{TLMM-NSAF}(\boldsymbol{w}(z))^T\hat{\boldsymbol{g}}_{TLMM-NSAF}(\boldsymbol{w}(z))\right]} \approx 2\frac{E\left[\dfrac{e_{i,D}^2}{\|\tilde{w}\|^2\|\tilde{x}\|^2}\right]}{E\left[\dfrac{e_{i,D}^2\|\tilde{x}\|^2}{\|\tilde{w}\|^4\|\tilde{x}\|^4}\right]}=2\|\tilde{w}\|^2\approx 2\left(\|h\|^2+\theta_i\right). \tag{A2}$$

## APPENDIX B

Using (40) we have:

$$\begin{aligned}
&E\left[\boldsymbol{r}(\boldsymbol{h})\boldsymbol{r}(\boldsymbol{h})^T\right]\\
&=\sum_{i=0}^{N-1}E\left[\frac{e_{i,D}^2\tilde{x}_i(z)\tilde{x}_i(z)^T}{\|\tilde{\boldsymbol{h}}\|^4\|\tilde{x}_i(z)\|^4}+\frac{e_{i,D}^3(z)\boldsymbol{h}\tilde{x}_i(z)^T}{\|\tilde{\boldsymbol{h}}\|^6\|\tilde{x}_i(z)\|^4}+\frac{e_{i,D}^3(z)\tilde{x}_i(z)\boldsymbol{h}^T}{\|\tilde{\boldsymbol{h}}\|^6\|\tilde{x}_i(z)\|^4}+\frac{e_{i,D}^4(z)\boldsymbol{h}\boldsymbol{h}^T}{\|\tilde{\boldsymbol{h}}\|^8\|\tilde{x}_i(z)\|^4}\right]\\
&=\frac{1}{\|\tilde{\boldsymbol{h}}\|^4}\sum_{i=0}^{N-1}\left\{\frac{E\left[e_{i,D}^2(z)\tilde{x}_i(z)\tilde{x}_i(z)^T\right]}{\|\tilde{x}_i(z)\|^4}+\frac{E\left[e_{i,D}^3(z)\boldsymbol{h}\tilde{x}_i(z)^T\right]}{\|\tilde{\boldsymbol{h}}\|^2\|\tilde{x}_i(z)\|^4}+\frac{E\left[e_{i,D}^3(z)\tilde{x}_i(z)\boldsymbol{h}^T\right]}{\|\tilde{\boldsymbol{h}}\|^2\|\tilde{x}_i(z)\|^4}+\frac{E\left[e_{i,D}^4(z)\right]\boldsymbol{h}\boldsymbol{h}^T}{\|\tilde{\boldsymbol{h}}\|^4\|\tilde{x}_i(z)\|^4}\right\}
\end{aligned} \tag{B1}$$

Using the properties of the kurtosis of the Gaussian distribution[5] and the assumptions 1 , it is easy to verify that

$$\begin{aligned}
E\left[e_{i,D}^2(z)\right]&=E\left[\left(v_i(z)-\boldsymbol{u}_i^T(z)\boldsymbol{h}\right)^2\right]\\
&=E\left[v_i(z)^2\right]+E\left[\left(\boldsymbol{u}_i^T(z)\boldsymbol{h}\right)^2\right]\\
&=\sigma_i^2\,\|\,\bar{\boldsymbol{h}}\,\|^2
\end{aligned} \tag{B2}$$

and

$$\begin{aligned}
E\left[e_{i,D}^4(z)\right]&=E\left[\left(v_i(z)-\boldsymbol{u}_i^T(z)\boldsymbol{h}\right)^4\right]\\
&=3\left\{E\left[\left(v_i(z)-\boldsymbol{u}_i^T(z)\boldsymbol{h}\right)^2\right]\right\}^2\\
&=3\sigma_i^4\,\|\,\bar{\boldsymbol{h}}\,\|^4
\end{aligned} \tag{B3}$$

Due to the mathematical definition of expectation,when $a$ and $b$ do not satisfy the independence condition, $E[ab]\neq E[a]E[b]$ . Then the expected term in B1 is calculated as:

$$\begin{aligned}
&E\left[e_{i,D}^2(z)\tilde{x}_i(z)\tilde{x}_i(z)^T\right]\\
&=E\Big[\left(v_n-\boldsymbol{u}_i^T(z)\boldsymbol{h}\right)^2\mathbf{R}+\left(v_n-\boldsymbol{u}_i^T(z)\boldsymbol{h}\right)^2\tilde{x}_i(z)\boldsymbol{u}_i^T(z)\\
&+\left(v_n-\boldsymbol{u}_i^T(z)\boldsymbol{h}\right)^2\boldsymbol{u}_i(z)\tilde{x}_i(z)^T\\
&+\left(v_n-\boldsymbol{u}_i^T(z)\boldsymbol{h}\right)^2\boldsymbol{u}_i(z)\boldsymbol{u}_i^T(z)\Big]\\
&=\sigma_{i,in}^2\,\|\,\tilde{\boldsymbol{h}}\,\|^2\left(\mathbf{R}+\sigma_{i,in}^2\boldsymbol{I}\right)
\end{aligned} \tag{B4}$$

and

$$
\begin{aligned}
&E\left[\left(v_i(z)-\boldsymbol{u}_i^T(z)\boldsymbol{h}\right)^3 \tilde{x}_i(z)\boldsymbol{h}^T\right] \\
&=E\Big[v_i(z)^3\,\boldsymbol{u}_i(z)\boldsymbol{h}^T-3v_i(z)^2\left(\boldsymbol{u}_i^T(z)\boldsymbol{h}\right)\boldsymbol{u}_i(z)\boldsymbol{h}^T \\
&+3v_i(z)\left(\boldsymbol{u}_i^T(z)\boldsymbol{h}\right)^2\boldsymbol{u}_i(z)\boldsymbol{h}^T-\left(\boldsymbol{u}_i^T(z)\boldsymbol{h}\right)^3\boldsymbol{u}_i(z)\boldsymbol{h}^T\Big] \\
&=-3\sigma_{i,in}^4\,\|\,\bar{\boldsymbol{h}}\,\|^2\,\boldsymbol{h}\boldsymbol{h}^T
\end{aligned}
\tag{B5}
$$

in the same way,we have

$$
E\left[\left(v_i(z)-\boldsymbol{u}_i^T(z)\boldsymbol{h}\right)^3 \boldsymbol{h}\tilde{x}_i(z)^T\right]=-3\sigma_{i,in}^4\,\|\,\bar{\boldsymbol{h}}\,\|^2\,\boldsymbol{h}\boldsymbol{h}^T \tag{B6}
$$

Where $\mathbf{R}=E\left[x_i(z)x_i^T(z)\right]$. Substituting (B3)–(B6) into (B1) results in (40).